# Superconductivity at 161 K in Thorium Hydride ThH$_{10}$: Synthesis and Properties

D. V. Semenok[1,&,*], A. G. Kvashnin[1,2,&], A. G. Ivanova[3], V. Svitlyk[4], V. Yu. Fominski[5], A.V. Sadakov[6], O.A. Sobolevskiy[6], V.M. Pudalov[6], I. A. Troyan[3] and A. R. Oganov[1,2,7,*]

[1] Skolkovo Institute of Science and Technology, Skolkovo Innovation Center 121025, 3 Nobel Street, Moscow, Russia

[2] Moscow Institute of Physics and Technology, 141700, 9 Institutsky lane, Dolgoprudny, Russia

[3] Shubnikov Institute of Crystallography, Federal Research Center Crystallography and Photonics, Russian Academy of Sciences, Moscow, 119333 Russia

[4] ID27 High Pressure Beamline, ESRF, BP220, 38043 Grenoble, France

[5] National Research Nuclear University MEPhI (Moscow Engineering Physics Institute), Kashirskoe sh., 31, Moscow 115409, Russia

[6] P.N. Lebedev Physical Institute, Russian Academy of Sciences, 119991 Moscow, Russia

[7] International Center for Materials Discovery, Northwestern Polytechnical University, Xi'an, 710072, China

**Corresponding Author**
*Dmitrii Semenok, E-mail: Dmitrii.Semenok@skoltech.ru
*Artem R. Oganov, E-mail: A.Oganov@skoltech.ru

## Highlights

- Superconductivity in *fcc*-ThH$_{10}$ at 159-161 K at the pressure 174 Gigapascals
- Very wide interval of stability of *fcc*-ThH$_{10}$ from 85 to 185 GPa.
- Upper critical magnetic field ThH$_{10}$ ~45 T.
- Novel discovered superhydride *hcp*-ThH$_9$ with $T_C$ of 146 K (170 GPa) and upper critical field ~38 T
- Newly discovered thorium hydrides: *I*4/*mmm*-ThH$_4$ and *Cmc*2$_1$-ThH$_6$

## Abstract

Here we report targeted high-pressure synthesis of two novel high-$T_C$ hydride superconductors, $P6_3/mmc$-ThH$_9$ and $Fm\bar{3}m$-ThH$_{10}$, with the experimental critical temperatures ($T_C$) of 146 K and 159-161 K and upper critical magnetic fields ($\mu H_C$) 38 and 45 Tesla at pressures 170-175 Gigapascals, respectively. Superconductivity was evidenced by the observation of zero resistance and a decrease of $T_C$ under external magnetic field up to 16 Tesla. This is one of the highest critical temperatures that has been achieved experimentally in any compounds, along with such materials as LaH$_{10}$, H$_3$S and HgBa$_2$Ca$_x$Cu$_2$O$_{6+z}$. Our experiments show that $fcc$-ThH$_{10}$ has stabilization pressure of 85 GPa, making this material unique among all known high-$T_C$ metal polyhydrides. Two recently predicted Th-H compounds, $I4/mmm$-ThH$_4$ (> 86 GPa) and $Cmc2_1$-ThH$_6$ (86-104 GPa), were also synthesized. Equations of state of obtained thorium polyhydrides were measured and found to perfectly agree with the theoretical calculations. New phases were examined theoretically and their electronic, phonon, and superconducting properties were calculated.

## Graphical Abstract

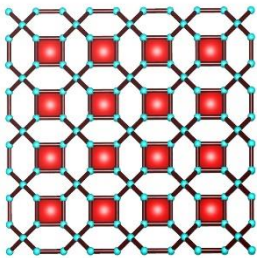
Computational prediction of new materials

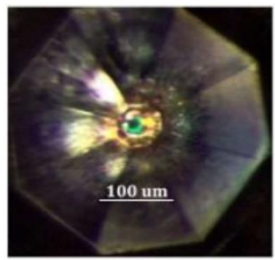
Targeted experimental high-pressure synthesis

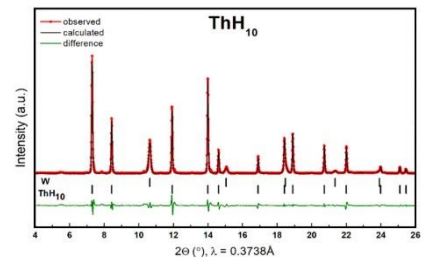
Experimental characterization of new materials

**Keywords:** thorium hydrides, USPEX, high pressure, superconductivity, X-ray diffraction

# 1. Introduction

Hydrogen is the first, simplest, and most common chemical element in the Universe. The transition of molecular hydrogen to the metallic atomic phase at high pressures was calculated for the first time in 1935 by Wigner and Huntington [1], with later estimations [2–4] showing that such metallization takes place only at very high pressures (450-550 GPa), which has not been technically achievable for a long time. In 2004, Ashcroft [5] proposed that the dissociation of the molecular hydrogen will take place at significantly lower pressures if it is chemically combined with other elements. Theoretical calculations [6,7] show that the stoichiometry (the amount of hydrogen in a compound) of stable metal hydrides in ambient conditions (e.g.,$Th_4H_{15}$) is not sufficient to identify the unique properties of the metallic hydrogen such as the high-$T_C$ superconductivity [8]. New chemical compounds possesses excess of hydrogen and unexpected structures and stoichiometries, which could be synthesized only at high pressures, are required to reach the high-$T_C$ superconductivity.

Room-temperature superconductivity has been an unattainable dream and subject of speculative discussions for a long time. However, the theoretical prediction of a high-temperature superconductor $H_3S$ [9] followed by the experimental confirmation [10,11] has opened a new field in the high-pressure physics devoted to investigations of superconducting hydrides. Recently predicted high-$T_C$ superconducting hydrides of thorium [12], actinium [13], lanthanum, and yttrium [14], and experimental proof of record high-$T_C$ superconductivity in $LaH_{10}$ at 250-260 K [15,16] and successful synthesis of previously predicted $CeH_9$ [17], $UH_7$, and $UH_8$ [18] motivated us to perform an experimental investigation of the Th-H system with a view to synthesize *fcc*-$ThH_{10}$, which was predicted to be a remarkable high-$T_C$ superconductor [12]. According to our previous predictions, it should be stable at pressures above 80 GPa [12] the lowest known stabilization pressure among the high-$T_C$ superconducting hydrides, and may have a $T_C$ up to 241 K and critical magnetic field $H_C$ = 71 T [12] at ~100 GPa, which makes the synthesis of this material very intriguing. Thus, the main goal of this study is the experimental synthesis of $Fm\overline{3}m$-$ThH_{10}$ as the most promising superconducting material [12]. The experimental high-temperature synthesis was performed by a short laser flash (200-500 ms) accompanied by the dissociation of ammonia borane, which leads to the generation of hydrogen simultaneously with the synthesis of higher hydrides.

# 2. Results and discussions

We prepared three samples, designated M1, M2, and M3. In this work we gradually increase the amount of hydrogen in the synthesized hydrides. In M2 sample, Th was loaded into a sublimated ammonia borane (AB) in a diamond anvil cell (DAC) and compressed to 98 GPa (see Table S2 in Supporting Information for details). The heating of M2 sample to 1800 K, performed at a single point by two laser pulses of 0.3 seconds with the power of 18 W, induces a pressure drop of 10 GPa which was measured by the Raman shift of diamond. The product was a mixture of *bct*-Th, *P*321-ThH$_4$, and *I*4*/mmm*-ThH$_4$ (Supporting Information Figure S1). Two obtained tetrahydrides were predicted previously [12]. It is interesting to note that at a pressure of 100 GPa the hydrogen content in the compound increased just by 5% compared to *c*-Th$_4$H$_{15}$, which is stable at normal conditions.

Due to unsatisfactory results of the first heating, the applied pressure was increased to 152 GPa. A subsequent heating to ~1400 K by five laser pulses of 0.3 seconds with the power of 50 W led to the decrease in pressure to 148 GPa. The products were metallic thorium and *I*4*/mmm*-ThH$_4$, as seen from the X-ray diffraction pattern (**Figure 1c**). Measurements taken at several points of the sample show only *I*4*/mmm*-ThH$_4$ with some *bcc*-W reflexes from the gasket (Supporting Information Figure S3). During an additional heating of M2 sample, we synthesized a new compound, *P*6$_3$/*mmc*-ThH$_9$, which was predicted previously [12] as a metastable phase. We performed a comprehensive theoretical and experimental analysis of its stability and superconducting properties.

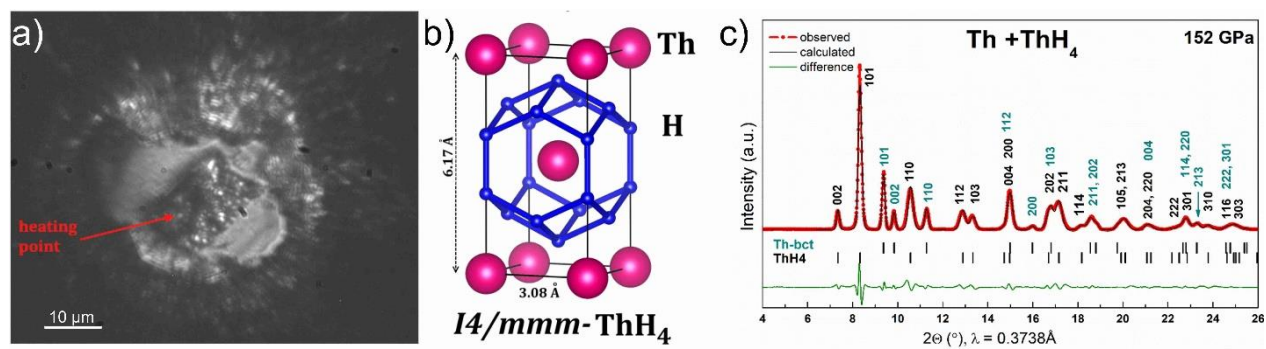

**Figure 1**. a) A photo of M2 sample after heating at 88 GPa. The heating point is shown by an arrow. b) Crystal structure of *I*4/*mmm*-ThH$_4$ at 90 GPa. c) The Le Bail refinement for *bct*-Th and *I*4/*mmm*-ThH$_4$ after the second heating at 152 GPa.

## 2.1. Synthesis of *P*6$_3$/*mmc*-ThH$_9$

One more step of laser heating of M2 was performed at 152 GPa by using four laser pulses of 0.3 seconds with the power of 60 W, which raised the temperature of M2 sample to ~2000 K at each flash. The XRD patterns of M2 sample after heating show the progressive formation of a new phase (**Figure 2**) at those points of the sample where only *I*4*/mmm*-ThH$_4$ was observed previously. New reflections were assigned to an unexpected hexagonal *P*6$_3$/*mmc*-ThH$_9$ phase, which forms during the ThH$_4$+2.5H$_2$→ThH$_9$ reaction. The details of its crystal structure and experimentally refined lattice parameters are shown in Supporting Information Tables S1 and S4, respectively. According to our predictions the thorium atoms here occupy the 2d Wyckoff position (1/3, 2/3, 3/4) in the hexagonal cell. The hydrogen atoms form an $H_{28}$ cage around each Th atom with the Th-H distance from 2.11 Å to 2.18 Å. Simultaneously with the formation of this phase, the width and asymmetry of $ThH_4$ peaks increased, possibly related to the $I4/mmm \rightarrow Fmmm$-$ThH_4$ phase transition (Figure S28b in Supporting Information). This orthorhombic modification of $ThH_4$ is found to be isenthalpic with $I4/mmm$-$ThH_4$ at this pressure, and to have a lower enthalpy at lower pressures.

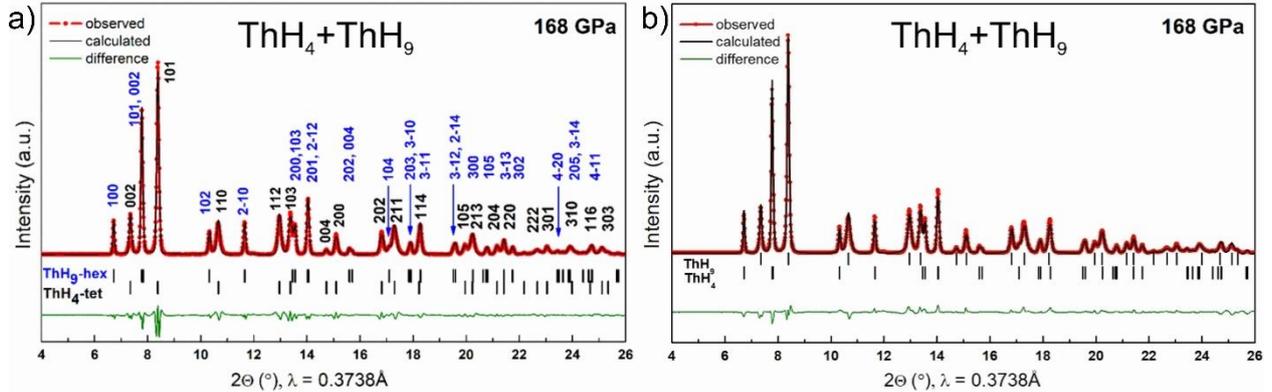

**Figure 2.** a) The Le Bail refinements and b) the Rietveld refinement of $P6_3/mmc$-$ThH_9$ and $ThH_4$ at 168 GPa after the fourth cycle of heating of M2 sample. The experimental data are shown in red, a fit approximation and residues are denoted by black and green lines, respectively. The R-factors (not corrected for background) for $ThH_4$+$ThH_9$ are: $R_p = 6.05\%$, $R_{wp} = 8.04\%$, $R_{exp} = 4.85\%$, $\chi_2 = 2.75\%$. The conventional Rietveld R-factors for $ThH_4$+$ThH_9$ are: $R_p = 0.9\%$, $R_{wp} = 12.8\%$, $R_{exp} = 7.68\%$, $\chi_2 = 2.75\%$.

To synthesize the higher thorium hydrides, we increased the pressure in M2 sample to 168 GPa and heated it by two laser pulses of 0.5 seconds with the power of 65 W, which raised the temperature to 2100 K at each flash. After the heating, the pressure in the sample rose by ~4 GPa. The XRD patterns show the absence of any phases other than $ThH_9$ and $ThH_4$ (**Figure 2**). However, the $ThH_4$/$ThH_9$ ratio changed during the heating and the amount of $ThH_9$ went up (see Figure S28 in Supporting Information). Thus, the increasingly monophase samples of $ThH_9$ were obtained by applying a series of laser heatings. $ThH_9$ was found to be isostructural with the previously predicted $P6_3/mmc$-$UH_9$ [18]. The shortest H-H distance in $ThH_9$ is smaller than that in $UH_9$ and equals 1.123 Å at 100 GPa and 1.099 Å at 150 GPa, compared to 1.144 Å at 100 GPa and 1.127 Å at 150 GPa for $UH_9$.

Our experiment with M2 sample showed that multiple laser heatings and pressure hikes increase the hydrogen content and lead to the formation of higher hydrides from the lower ones using $NH_3BH_3$ as a source of hydrogen. However, the first step of laser heating performed at the relatively low pressure of ~88 GPa precluded the synthesis of thorium superhydrides because of the formation of a mixture of lower hydrides, mainly $I4/mmm$-$ThH_4$. We also found that overheating the samples above 2200 K led to a destruction of the diamond anvil cell, while heating to lower than 1400 K was inefficient and resulted in mixtures of hydrides with a low H-content.

2.2. Stability and physical properties of $P6_3/mmc$-ThH$_9$

$P6_3/mmc$-ThH$_9$ was found to be a thermodynamically metastable phase in our previous calculations, therefore, we did not study its physical properties [12]. Having observed this phase experimentally, here we calculate its zero-point energy as $U_{ZPE}(V,T) = 1/2 \int g(\omega(V))\hbar\omega d\omega$ and find that it leads to the stabilization of ThH$_9$ (**Figure 3a**). A similar stabilization was found for the recently studied $C2/m$-Ca$_2$H$_5$ [19].

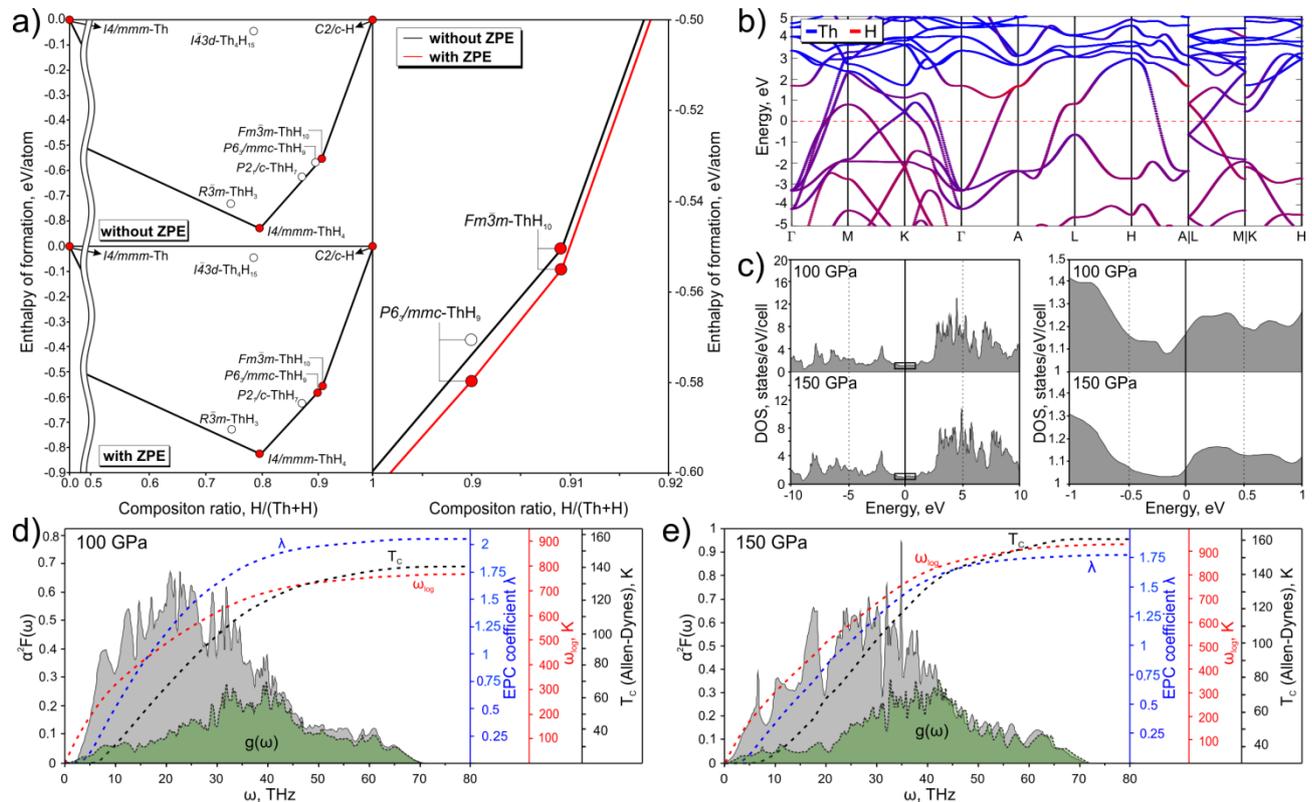

**Figure 3**. a) (left) A thermodynamic convex hull of the Th-H system with and without the zero-point energy (ZPE) contribution at 150 GPa. (right) A magnified region near ThH$_9$ and ThH$_{10}$ showing that ThH$_9$ is stabilized by the zero-point energy. b) Band structure of $P6_3/mmc$-ThH$_9$ at 150 GPa. Hydrogen and thorium are shown in red and blue, respectively. c) The electronic DOS of $P6_3/mmc$-ThH$_9$ (Z = 2) at 100 GPa and 150 GPa computed for energy ranges ±10 eV and ±1 eV around the Fermi level. d), e) The phonon density of states (shown in green, in a.u.), the Eliashberg $\alpha^2F(\omega)$ function (in black), and the superconducting parameters of $P6_3/mmc$-ThH$_9$ at 100 GPa and 150 GPa.

The crystal structure of $P6_3/mmc$-ThH$_9$ is similar to that of the previously predicted and synthesized CeH$_9$ [17] which can be explained by the similarity of electronic structures of Ce ($4f^15d^1$) and Th ($6d^2$) atoms. Zero-point energy plays a crucial role in stabilizing ThH$_9$ at pressures above 105 GPa (**Figure 3a**). This lower limit of thermodynamic stability of ThH$_9$ perfectly agrees with our experimental data. The study of the dynamic stability of ThH₉ at different pressures shows the absence of imaginary phonon frequencies at 100 GPa and 150 GPa (**Figure 3d,e** and Supporting Information Figure S21), which is consistent with our experiments. At pressures below 100 GPa, ThH₉ becomes dynamically unstable. Under the studied conditions, all thorium hydrides are metallic in both experiment and theory. The calculated band structure at 150 GPa shows a predominant contribution of hydrogen in the energy range close to the Fermi level (red color in **Figure 3b**), which lies in an almost flat "valley" of the density of states.

We calculated the Eliashberg $\alpha^2F(\omega)$ function at 100 GPa and 150 GPa to find the critical temperature ($T_C$) and critical magnetic field ($H_C$), shown in **Figure 3d,e**. At 150 GPa, ThH₉ has a high electron-phonon coupling (EPC) coefficient $\lambda = 1.73$ and a high logarithmic frequency $\omega_{log} = 957$ K, giving the predicted $T_C = 123\text{-}145$ K. Calculated superconducting parameters are summarized in **Table 1**.

We estimated the critical magnetic field and the jump in specific heat for ThH₉ (see **Table 1**). The calculation of the upper critical magnetic field gives $\mu_0H_C(0) \sim 39$ T at 150 GPa. The estimated coherence length $\xi_{BCS} = 0.5\sqrt{h/\pi e\, H_C}$ is 30 Å. The isotopic coefficient for $P6_3/mmc$-ThH₉ at 150 GPa is 0.48 using $\mu^* = 0.1$ for Allen-Dynes (A-D) equation (**Table 1**) in the harmonic approximation, and $T_C$(ThD₉) is 89–104 K. $T_C$ of ThH₉ slightly increases with pressure ($dT_C/dP = +0.06\text{–}0.1$ K/GPa), while for $Fm\bar{3}m$-ThH₁₀, the opposite dependence was predicted [12]. Extrapolation of this data to experimentally studied pressure point – 170 GPa leads to predicted interval of 125-147 K for $T_C$ and 32-36 T for $\mu_0H_C(0)$ which coincides perfectly with the experiment (see Section 2.4).

**Table 1.** Calculated parameters of the superconducting state of $P6_3/mmc$-ThH₉ at 100 GPa and 150 GPa using Allen-Dynes (A-D) formula. Here $\gamma$ is the Sommerfeld constant, the values are given at $\mu^*$ equal to 0.15-0.1, $\beta$ is the isotopic coefficient.

| Parameter | $P6_3/mmc$-ThH₉ | |
|---|---|---|
| | 100 GPa | 150 GPa |
| $\lambda$ | 2.15 | 1.73 |
| $\omega_{log}$, K | 728 | 957 |
| $\beta$ | 0.47-0.48 | 0.47-0.48 |
| $T_C$ (A-D), K | 118-138 | 123-145 |
| $T_C$ ($P6_3/mmc$-ThD₉), K | 85-99 | 89-104 |
| $\Delta(0)$, meV | 31.2-35.2 | 29.6-33.9 |
| $\mu_0H_C(0)$, T | 37-41 | 33-37 |
| $\Delta C/T_C$, mJ/mol·K² | 25.8 | 19.3-19.9 |
| $\gamma$, mJ/mol·K² | 8.69 | 7.52 |
| $R_\Delta = 2\Delta(0)/k_BT_C$ | 5.11-5.24 | 4.74-4.89 |

## 2.3. Synthesis of $Fm\bar{3}m$-ThH$_{10}$ and $Fm\bar{3}m \rightarrow Immm$ phase transition

ThH$_{10}$ was predicted to be stable and have an extremely high $T_C$ [12]. Here we used M3 sample (Supporting Information Table S2), with the increased initial pressure of 170 GPa, heating it to 1800 K by four laser pulses of 0.3 seconds each, which induced an increase in pressure to 183 GPa. The analysis of the measured X-ray diffraction pattern of the entire sample area showed that the sample consisted of only $Fm\bar{3}m$-ThH$_{10}$ (**Figure 4a**), theoretically predicted and studied earlier [12]. The XRD pattern also contained reflections from the tungsten gasket because of the relatively small size of the sample and chamber, ~20 μm. By the subsequent stepwise reduction of pressure in M3 sample from 183 GPa to 85 GPa (**Figure 4b**), we measured the equation of state of $Fm\bar{3}m$-ThH$_{10}$ (**Figure 6**) and found the lower limit of stability of this phase. The experimental lattice parameters of this phase are shown in Table S5 (see Supporting Information). At a pressure close to the theoretically predicted lower limit of its stability (85 GPa [12]), we observed broadening and splitting of ThH$_{10}$ XRD peaks accompanying the $Fm\bar{3}m \rightarrow Immm$-ThH$_{10}$ phase transition (see Supporting Information, Figure S27). $Immm$-ThH$_{10}$ ($a = 5.304(2)$ Å, $b = 3.287(1)$ Å, $c = 3.647(2)$ Å, $V = 74.03$ Å$^3$) is the closest in energy metastable modification of ThH$_{10}$ that appears in experiments at pressures below 100 GPa. An additional signal observed at $2\theta = 7.455°$ was caused by the decomposition of ThH$_{10}$ producing $P6_3/mmc$-ThH$_9$.

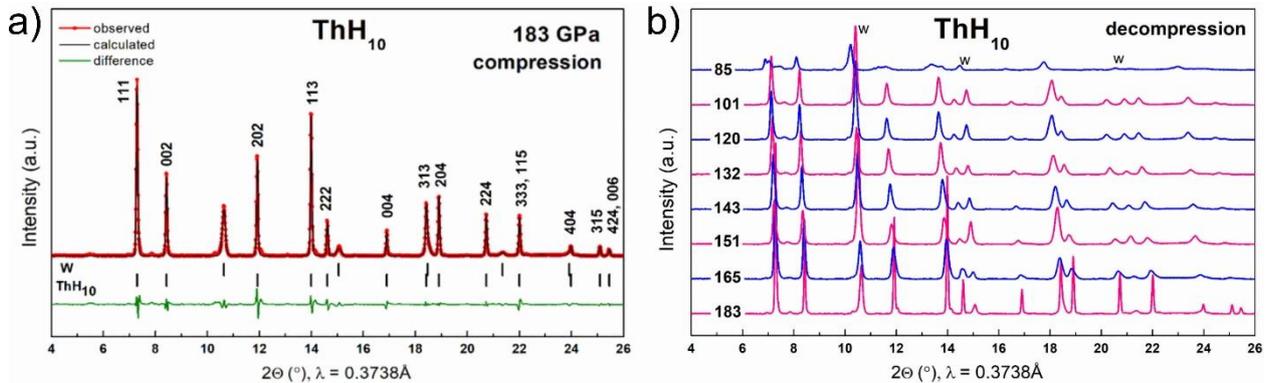

**Figure 4.** a) The Le Bail refinement of $Fm\bar{3}m$-ThH$_{10}$ and $bcc$-W at 183 GPa. The experimental data are shown in red, a fit approximation and residues are denoted by black and green lines, respectively. The Rietveld refinement for ThH$_{10}$ is given in Figure S19 (Supporting Information). b) Experimental XRD patterns of $Fm\bar{3}m$-ThH$_{10}$ in the pressure range of 183-85 GPa.

The cubic hydrogen-rich phase $Fm\bar{3}m$-ThH$_{10}$ was the main goal of this study due to its predicted superconductivity with $T_C$ up to 241 K [12] at 90-100 GPa. This is also in agreement with phonon calculations, which prove its dynamic stability down to ~85 GPa (Figure S25 in Supporting Information). ThH$_{10}$ is structurally similar to the already predicted cubic decahydrides ScH$_{10}$ [19,20], YH$_{10}$ [14], LaH$_{10}$ [15,16,22], and AcH$_{10}$ [13]. The shortest H-H distances in these materials form the sequence $d_{min}$(AcH$_{10}$) < $d_{min}$(ThH$_{10}$) < $d_{min}$(LaH$_{10}$) at 150 GPa (1.07 Å < 1.119 Å < 1.164 Å).

## 2.4. Measurements of superconducting properties of ThH$_{10}$ and ThH$_9$

To measure the temperature of the transition of thorium hydride to the superconducting state, a cell M1 with diamond anvils and Ta/Au electrodes was prepared. We used the anvils with a 50 μm central culet beveled to 300 μm at 8.5°. Four Ta electrodes (~200 nm) with gold plating (~80 nm) were sputtered onto the piston diamond. A composite gasket consisting of a tungsten ring and a MgO epoxy mixture insert was used to isolate the electrical leads.

An ~1 μm-thick thorium sample was sandwiched between the electrodes and AB in the gasket hole 20 μm in diameter. The pressure in the cell was increased to 179 GPa. The heating of the sample was performed by pulses of a Nd:YAG infrared laser with the wavelength λ = 1.064 μm, power of 35–40 W, and duration of 400 ms. The pressure in the chamber reduced to 174 GPa after the laser heating (Figure S29 in Supporting Information). The temperature dependence of the resistance is shown in **Figure 5**. After the measurement, the cell was transferred to a synchrotron to confirm the structure of the hydride. The Le Bail refinement for the sample is shown in Figure S30a (Supporting Information).

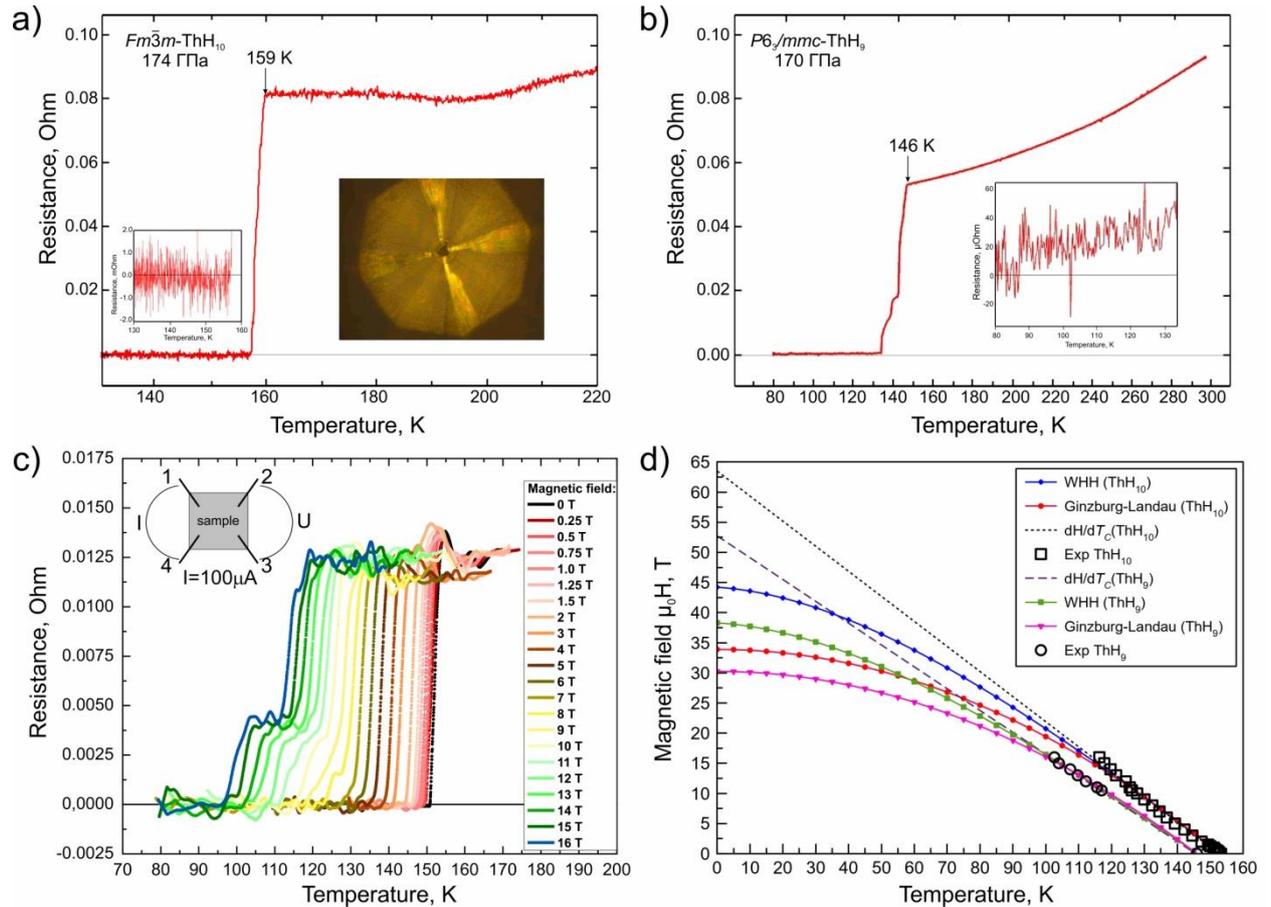

**Figure 5**. Observation of superconductivity in a) ThH$_{10}$ and b) ThH$_9$. The temperature dependence of the resistance (*R*) of thorium superhydride was determined in a sample synthesized from Th+NH$_3$BH$_3$. The resistance was measured with four electrodes deposited on a diamond anvil on which the sample was placed (the right panel inset) at excitation current 100μA. The resistance near the zero point is shown in a smaller scale in the insets; c) dependence of resistance on temperature under an external magnetic field at 170 GPa; d) dependence of critical temperature ($T_C$) of ThH$_{10}$ and ThH$_9$ on magnetic field.

We found that depending on the synthesis pressure, two different groups of superconducting transitions with $T_C$ of 159-161 K (ThH$_{10}$, **Figure 5a** and Figure S30b in Supporting Information) and 146 K (ThH$_9$, **Figure 5b**) can be observed, where the electrical resistance decreased sharply to zero (from 60 mΩ to 1-10 μΩ). The superconducting nature of the transitions was verified by its dependence on the external magnetic fields in the range 0-16 Tesla. **Figure 5c** shows that an applied magnetic field of $\mu_0 H$ = 4.1 T reduces the onset of the superconducting transition by 10 K. Starting from 8 T a new resistance step appears, **Figure 5c**. This step may be well explained by impurities of *hcp*-ThH$_9$ which has lower $H_C$ and $|dH/dT_C|$ = 0.36 T/K, and cannot be distinguished from the transition in *fcc*-ThH$_{10}$ in weaker magnetic fields due to proximity-induced superconductivity [23].

The extrapolation of the temperature-dependent upper critical fields $\mu_0 H_{C2}(T)$ by linear function (dotted and dashed lines in **Figure 5d**) yields about 64 and 53 T for *fcc*-ThH$_{10}$ and *hcp*-ThH$_9$, respectively. The use of Werthamer-Helfand-Hohenberg (WHH) model [24] simplified by Baumgartner et al. [25] gives the values of 45 T (ThH$_{10}$) and 38 T (ThH$_9$) (**Figure 5d**).

It is interesting to compare the experimentally obtained values with theoretical calculations based on the Bardeen–Cooper–Schrieffer and Migdal–Eliashberg theories. Superconductivity of thorium decahydride in the harmonic approximation was studied before [12] and the critical temperature was found to be in the range 150–183 K at 200 GPa (the Allen-Dynes formula [26], $\mu^*$ = 0.15-0.1). A linear extrapolation to the lower pressure of 174 GPa gives $T_C$ in the range 160-193 K in a good agreement with measured interval 159-161 K.

Calculations of the electron-phonon interaction in cubic ThH$_{10}$ at 174 GPa yield λ = 1.75, $\omega_{log}$ = 1520 K, $T_C$ ~ 160 K, and $\mu_0 H_{C2}(0)$ = 38 T (**Table 1**), which is slightly lower than experimentally found value (45 T, WHH model, **Figure 5d**), but much lower than the Clogston-Chandrasekhar paramagnetic limit (when magnetostatic polarization energy exceeds Cooper pairs condensation energy) $\Delta_0/\sqrt{2}$ (359 T) [27]. Estimation of the average Fermi velocity $V_F$ ~ 4.12×10$^5$ m/s makes it possible to calculate the London penetration depth $\lambda_L$ ~ 136 nm, coherence length $\xi_{BCS}$ = 29 nm, and lower critical magnetic field $\mu_0 H_{C1}$ = 0.024 T. The critical current density ($J_C = e n_e V_L$), evaluated by Landau criterion for superfluidity [28] $V_L = min \frac{\varepsilon(p)}{p} \cong \frac{\Delta_0}{\hbar k_F}$, was calculated to be ~ 3.16×10$^8$ A/cm$^2$, much higher than in H$_3$S [10]. The calculated Ginzburg-Landau parameter [29] is over 46 which is typical for II type superconductors.

## 2.5. Equation of states

By decompressing M2 sample from 168 GPa to 86 GPa and further down until its destruction (cracking, Supporting Information Figure S4a), we studied the equation of state of $I4/mmm$-ThH$_4$ and $P6_3/mmc$-ThH$_9$ and obtained the previously predicted low-symmetry thorium hexahydride $Cmc2_1$-ThH$_6$ [12] (Supporting Information Figure S4). The Le Bail refinements of observed hydrides are shown in Figures S7-S9 in Supporting Information. The experimental equations of state (**Figure 6**) of the synthesized thorium hydrides are in close agreement with theoretical predictions [12].

The experiments in this study can be used to assess the predictive power of the evolutionary algorithm USPEX [30–32], which was used to predict thorium polyhydrides [12]. Here, we synthesized thorium decahydride ThH$_{10}$ along with thorium hexa- and tetrahydrides ($Cmc2_1$-ThH$_6$, $P321$- and $I4/mmm$-ThH$_4$), which were predicted before [12]. Also, $hcp$-ThH$_9$, a new superconductor with $T_C$ of 146 K, was synthesized. Thus, the experiment proves a good reliability of the USPEX predictions.

The example of the Th-H system, together with the Ce-H, U-H and La-H systems, shows that the lower bound of stability of metal polyhydrides shifts toward lower pressures when going down the Periodic Table. Increasing the complexity to ternary (A-B-H) systems might lead to further lowering of the synthesis pressure and increase $T_C$. Theory is set to play a leading role in the study of such systems.

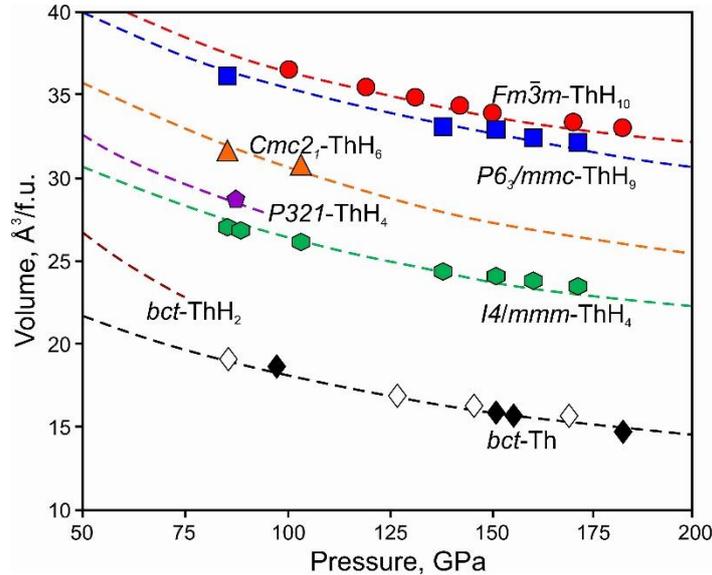

**Figure 6**. Equations of state of polyhydrides synthesized in this work in comparison with theoretical predictions and previously published data from Ref. [35].

## 3. Conclusions

The high-$T_C$ superconductors $Fm\bar{3}m$-ThH$_{10}$ and new $P6_3/mmc$-ThH$_9$, as well as three other previously predicted polyhydrides, $Cmc2_1$-ThH$_6$, $P321$-ThH$_4$, and $I4/mmm$-ThH$_4$ were synthesized, confirming earlier theoretical predictions [12]. Predicted lower bound of dynamical stability of ThH$_{10}$ (80 GPa) perfectly agrees with the experimentally determined pressure of ~85 GPa, i.e., *fcc*-ThH$_{10}$ has a unique combination of high superconducting $T_C$ (159-161 K at 175 GPa) critical magnetic field $\mu_0H_C(0)$ of 45 Tesla, and the stability pressure, which is much lower than for other high-$T_C$ hydrides. The measured parameters of the superconducting state of *hcp*-ThH$_9$ show $T_C$ of 146 K, the upper critical magnetic field $\mu_0H_C(0)$ of 38 Tesla, and the superconducting gap of ~35 meV. In addition, we experimentally observed pressure-driven phase transitions in ThH$_4$ ($P321 \rightarrow I4/mmm \rightarrow Fmmm$) and ThH$_{10}$ ($Fm\bar{3}m \rightarrow Immm$). The rich observed chemistry of the Th-H system agrees with our theoretical predictions. Stable and metastable phases predicted using the USPEX method can serve as a good guide for an experimental synthesis. The experiments performed will have a strong impact on our understanding of high-pressure chemistry of metal hydrides, bringing us closer to attaining room-temperature superconductivity at high and, hopefully, ambient pressures.

## 4. Methods

### 4.1. Experiment

To perform this experimental study, three diamond anvil cells (DACs) were loaded. The diameter of the working surface of diamond anvils was 280 μm, they were beveled at an angle of 8.5° to form a culet of 50 μm. The data on these DACs are shown in Supporting Information Table S2. X-ray diffraction patterns of all samples in diamond anvil cells were recorded at ID27 synchrotron beamline at the European Synchrotron Radiation Facility (Grenoble, France) with the use of a focused (1.7 × 2.3 μm) monochromatic X-ray beam of 33 eV (λ = 0.3738 Å) and a Perkin-Elmer area detector placed at a distance of 364.12 mm from the sample. The exposure time was 30–100 s. CeO$_2$ standard was used for the distance calibration. The X-ray diffraction data were analyzed and integrated using Dioptas software package (version 0.4) [30]. The full profile analysis of the diffraction patterns and the calculation of the unit cell parameters were performed in JANA2006 program [31] using the Le Bail method. [32] The heating of the samples was done by pulses of Nd:YAG infrared laser with the wavelength λ = 1 μm, power of 18-40 W, and duration of 300-500 ms. Temperature measurements were carried out using the decay curve of black body radiation within the Planck and Vino formulas at the laser heating system of ID27 beamline ESRF. The exact position of the heating spot was determined by an optical flash of a heated material detected by a CCD camera. The applied pressure was measured by the edge position of the Raman signal of diamond [33] using Acton SP2500 spectrometer with PIXIS:100 spectroscopic-format CCD. The pressure measurements were made using the Raman signals of a DAC. In the experimental X-ray images, the reflections from the BNH compounds are absent because of their amorphous state and weak X-ray scattering from light atoms.

### 4.2. Theory

The equations of state of the predicted ThH$_4$, ThH$_6$, ThH$_9$ and ThH$_{10}$ phases were calcualted using density functional theory (DFT) [40,41] within the generalized gradient approximation (Perdew-Burke-Ernzerhof functional) [42], and the projector-augmented wave method [43,44] as implemented in the VASP code [45–47]. Plane wave kinetic energy cutoff was set to 500 eV and the Brillouin zone was sampled using Γ-centered $k$-points meshes with resolution $2\pi \times 0.05$ Å$^{-1}$. Obtained dependences of the volume on pressure were fitted by the 3$^{rd}$ order Birch-Murnaghan equation [48] to determine the main parameters of the EOS, namely $V_0$, $K_0$ and $K'$, where $V_0$ is the equilibrium volume, $K_0$ is the zero-pressure bulk modulus and $K'_0$ its pressure derivative of the bulk modulus. Fitting was done using the EOSfit7 program [49]. We also calculated phonon densities of states of studied materials using the finite-displacements method (VASP and PHONOPY [50,51]).

Calculations of superconducting $T_C$ were carried out using QUANTUM ESPRESSO (QE) package [52]. Phonon frequencies and electron-phonon coupling (EPC) coefficients were computed using density-functional perturbation theory [53], employing plane-wave pseudopotential method and Perdew-Burke-Ernzerhof exchange-correlation functional [42]. In our *ab initio* calculations of the electron-phonon coupling (EPC) parameter $\lambda$, the first Brillouin zone was sampled using 4×4×4 $q$-points mesh, and a denser 24×24×24 $k$-points mesh (with Gaussian smearing and σ = 0.025 Ry, which approximates the zero-width limits in the calculation of $\lambda$). $T_C$ was calculated from the Eliashberg equations [54] which were solved by iterative self-consistent method for the imaginary part of the order parameter $\Delta(T, \omega)$ (superconducting gap) and the renormalization wave function $Z(T, \omega)$ [55] (see Supporting Information). More approximate estimates of $T_C$ were made using the Allen-Dynes formula [26].

Our calculations show that all found thorium hydrides are diamagnetic and are thermodynamically stable at conditions where experiments found them. As the new unexpected phase $P6_3/mmc$-ThH$_9$ has not been studied before, we performed additional calculations of its electronic, phonon and superconducting properties in the pressure range 100-200 GPa.

## 5. Author Contributions:

[&] These authors contributed equally to this work.
I.A.T, D.V.S., A.G.I., V.S., V.Yu.F. performed the experiment, A.G.K. and A.R.O. prepared theoretical analysis.

# 6. Acknowledgments

The work was performed on ESRF (Grenoble, France), station ID27. The authors express their gratitude to K. German (IGIC) and I. Polovov (URFU) for the technical assistance. The work on high-pressure experiments was supported by the Ministry of Science and Higher Education of the Russian Federation within the State assignment of the FSRC "Crystallography and Photonics" of RAS and by the Russian Science Foundation (Project No.19-12-00414). The authors thank the RFBR foundation project №19-03-00100. A.G.K. thanks the FACIE foundation UMNIK grant №13408GU/2018 for the financial support of this work. A.R.O. thanks the Russian Science Foundation (grant 19-72-30043).

# Supporting Information

# Superconductivity at 161 K in Thorium Hydride ThH$_{10}$: Synthesis and Properties

D. V. Semenok[1,&,*], A. G. Kvashnin[1,2,&], A. G. Ivanova[3], V. Svitlyk[4], V. Yu. Fominski[5], A.V. Sadakov[6], O.A. Sobolevskiy[6], V.M. Pudalov[6], I. A. Troyan[3] and A. R. Oganov[1,2,7,*]

[1] Skolkovo Institute of Science and Technology, Skolkovo Innovation Center 121025, 3 Nobel Street, Moscow, Russia

[2] Moscow Institute of Physics and Technology, 141700, 9 Institutsky lane, Dolgoprudny, Russia

[3] Shubnikov Institute of Crystallography, Federal Research Center Crystallography and Photonics, Russian Academy of Sciences, Moscow, 119333 Russia

[4] ID27 High Pressure Beamline, ESRF, BP220, 38043 Grenoble, France

[5] National Research Nuclear University MEPhI (Moscow Engineering Physics Institute), Kashirskoe sh., 31, Moscow 115409, Russia

[6] P.N. Lebedev Physical Institute, Russian Academy of Sciences, 119991 Moscow, Russia

[7] International Center for Materials Discovery, Northwestern Polytechnical University, Xi'an, 710072, China

**Corresponding Authors**
*Dmitrii Semenok, e-mail: Dmitrii.Semenok@skoltech.ru
*Artem R. Oganov, e-mail: A.Oganov@skoltech.ru

**Author Contributions:**
[&] These authors contributed equally to this work.
D.V.S., I.A.T, and A.G.I. performed the experiment. V.Y.F. and I.A.T. deposited electrodes and performed superconductivity measurements. A.G.K. and A.R.O. prepared the theoretical analysis. D.V.S., A.G.K. and A.G.I. contributed to the interpretation of the results. D.V.S., A.G.K., A.R.O. wrote the manuscript. All authors provided critical feedback and helped shape the research, analysis and manuscript.

# Contents

# Structures of newly discovered *P6₃/mmc*-ThH₉, *Cmc2₁*-ThH₆, and *P321*-ThH₄

**Table S1.** Crystal structures of new Th-H phases.

| Phase | Predicted stability range, GPa | Lattice Parameters at 100 GPa | Coordinates | | | |
|---|---|---|---|---|---|---|
| *P6₃/mmc*-ThH₉ 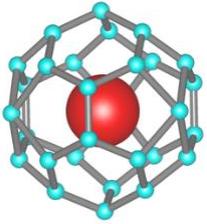 | 100-200 | a = 3.799 Å<br>c = 5.678 Å | Th1 | 0.333 | 0.667 | 0.750 |
| | | | H1 | 0.156 | 0.312 | 0.061 |
| | | | H2 | 0.333 | 0.667 | 0.348 |
| | | | H3 | 0.000 | 0.000 | 0.25 |
| *Cmc2₁*-ThH₆ 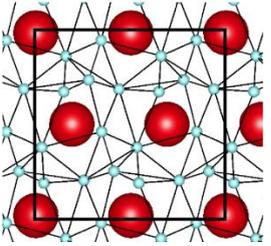 | 25-90 | a = 3.65 Å<br>b = 6.038 Å<br>c = 5.834 Å | Th1 | 0.000 | -0.345 | -0.345 |
| | | | H1 | 0.000 | -0.066 | 0.397 |
| | | | H2 | 0.000 | 0.001 | 0.114 |
| | | | H3 | 0.000 | -0.313 | 0.028 |
| | | | H4 | 0.252 | 0.092 | 0.347 |
| | | | H5 | 0.356 | 0.197 | -0.166 |
| *P321*-ThH₄ 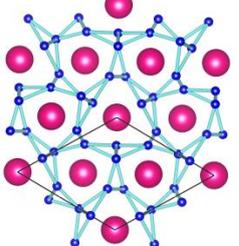 | 15-85 | a = 5.44 Å<br>c = 3.24 Å | Th1 | 0.000 | 0.000 | 0.500 |
| | | | Th2 | 0.333 | 0.667 | 0.185 |
| | | | H1 | 0.369 | 0.000 | .500 |
| | | | H2 | 0.251 | 0.000 | 0.000 |
| | | | H3 | -0.402 | -0.082 | -0.283 |

# Details of the experiment

In DAC M1, pure tantalum was used for electrodes. It can also form hydrides, which should be taken into account. However, the comparison of simulated XRD patterns for TaH$_6$, TaH$_4$ and TaH$_3$ from Ref. [1] with our data does not show the presence of these compounds in the samples.

For the *in situ* production of hydrogen we use the sublimated ammonia borane (99.9%). This compound decomposes stepwise as follows: [2]

$$NH_3BH_3 \rightarrow (NH_2BH_2)_n + H_2 \rightarrow (NHBH)_n + H_2 \rightarrow c\text{-}BN + H_2 \text{ (over 1440 K)}$$

This approach was successfully applied during the synthesis of LaH$_{10}$ [3] and has certain advantages compared to the loading of hydrogen. The reduction of the working volume of the anvil cell and culet easily leads to a leakage of hydrogen which subsequently reacts with the metal target, forming undesirable lower hydrides at room temperature.

**Table S2.** Experimental parameters of DACs. The culet diameter was 50 μm, the gasket thickness was 12 μm.

| #cell | Initial pressure, GPa | Gasket | Sample size, μm | Composition/load |
|---|---|---|---|---|
| M1 | 101 | MgO/epoxy | 30 | Th/BH$_3$NH$_3$ + Au/Ta electrodes* |
| M2 | 98 | W | 30 | Th/BH$_3$NH$_3$ |
| M3 | 170 | W | 10 | Th/BH$_3$NH$_3$ |

*The thickness of the deposited Ta and Au layers was 70 nm and 30 nm, respectively.

# Synthesis of *I4/mmm*-ThH$_4$

The X-ray diffraction pattern of M2 sample before heating is shown in Figure S12, it is *I4/mmm*-Th phase with lattice parameters $a$ = 2.903(1) Å, $c$ = 4.421(1) Å. The heating of M2 sample to 1800 K (see Details of the experiments) performed at a single point by two laser pulses of 0.3 seconds with the power of 18 W is shown in **Figure S1**a. The shape of the sample did not change, while the optical flash and changes in the surface properties were observed through a microscope (**Figure S1**a). The appearance of the spotty powder diffraction rings after the heating indicates melting and recrystallization followed by the formation of relatively large crystallites of a new phase. The pressure measured by the Raman shift of diamond was 88 GPa, which is ~10 GPa lower than the initial pressure in the cell.

As a result, we obtained a complex mixture of lower thorium hydrides with at least 3 phases: *P*321-ThH$_4$ ($a$ = 5.500(1) Å, $c$ = 3.290(1) Å, $V$ = 86.18(3) Å$^3$), [4] *bct*-Th ($a$ = 3.042(1) Å, $c$ = 4.394(2) Å, $V$ = 40.67(1) Å$^3$), and *I4/mmm*-ThH$_4$ ($a$ = 3.058(1) Å, $c$ = 6.120(1) Å, $V$ = 57.23(2) Å$^3$, Z = 2). *I4/mmm*-ThH$_4$ is a new compound where each Th atom is coordinated by 18 hydrogen atoms forming an H$_{18}$ cage (**Figure S1b**). The X-ray diffraction pattern of the synthesized phase is shown in **Figure S1c**. *P*321-ThH$_4$ has a trigonal structure with two symmetrically inequivalent thorium atoms in *1b* and *2d* Wyckoff positions, where each Th atom is coordinated by 15 hydrogen atoms. The crystal structure data is summarized in Table S1.

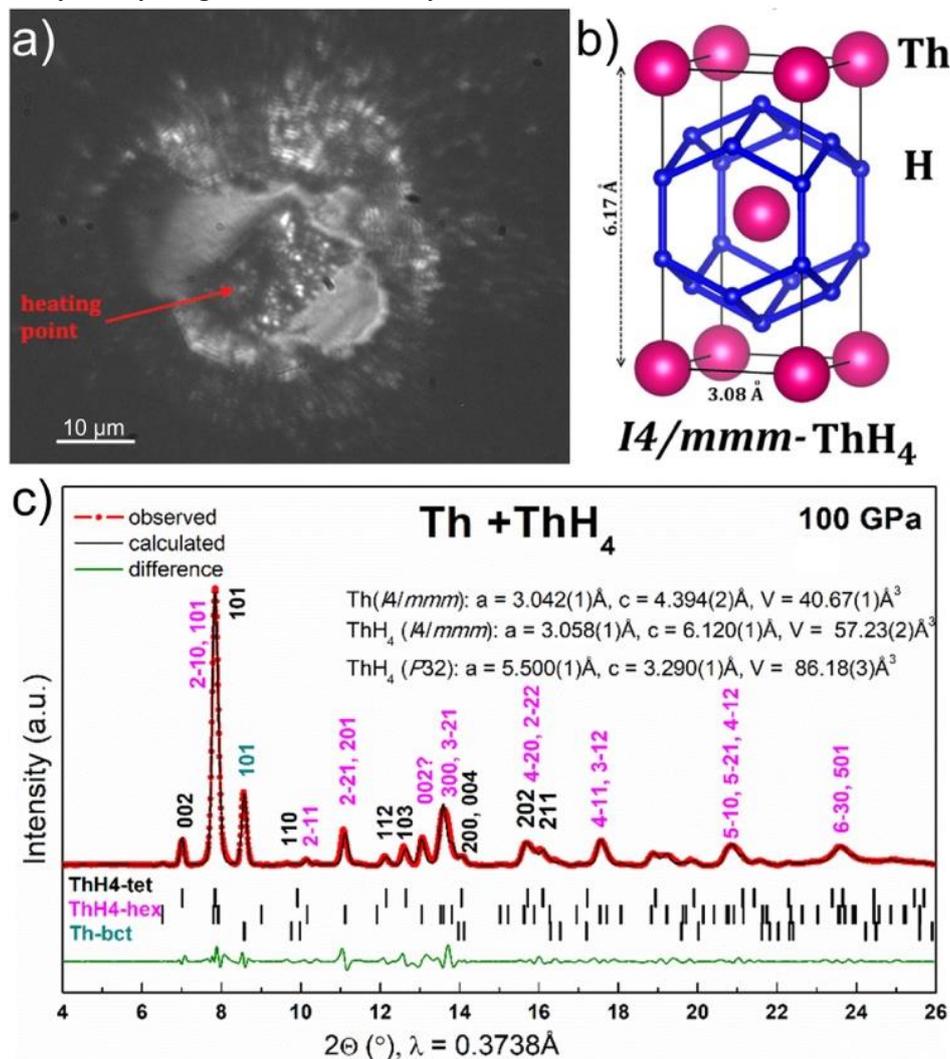

**Figure S1**. (a) A photo of M2 sample after heating at 88 GPa. The heating point is shown by an arrow. (b) Crystal structure of *I4/mmm*-ThH$_4$ at 90 GPa. (c) The Le Bail refinement for *I4/mmm*-ThH$_4$, *P*321-ThH$_4$, and *bct*-Th with the experimentally obtained lattice parameters at 88 GPa (Run 1).

## *P*321-ThH$_4$ → *I*4/*mmm*-ThH$_4$ Transformation

The lattice parameters of the new phase are shown in **Table S3**. Thus, the only result of compression to over 50 GPa and heating was the *P*321 → *I*4/*mmm* phase transition, which indicates an extremely high stability of thorium tetrahydride.

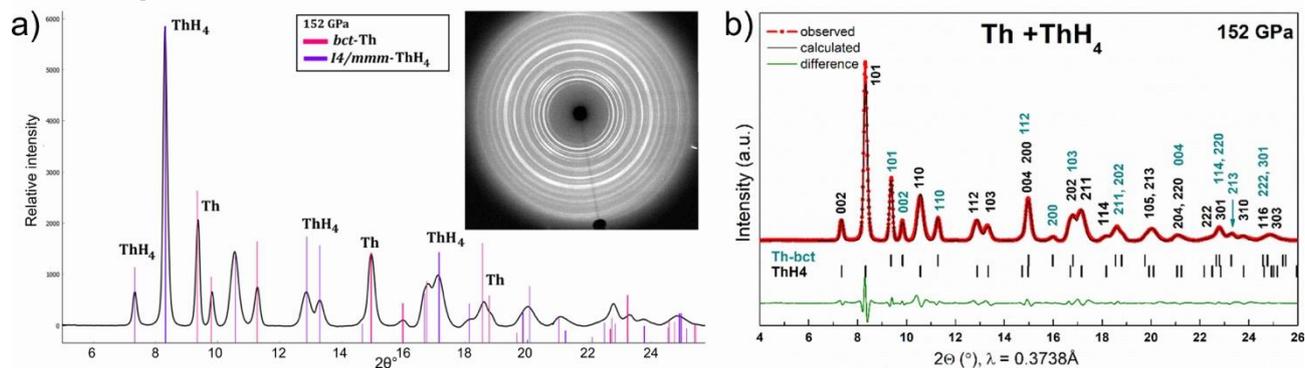

**Figure S2**. (a) XRD pattern of M2 sample after the second heating at 152 GPa (Run 2) in comparison with simulated *I*4/*mmm*-ThH$_4$ and *bct*-Th. (b) The Le Bail refinement for *bct*-Th and *I*4/*mmm*-ThH$_4$ after the second heating at 152 GPa (Run 2).

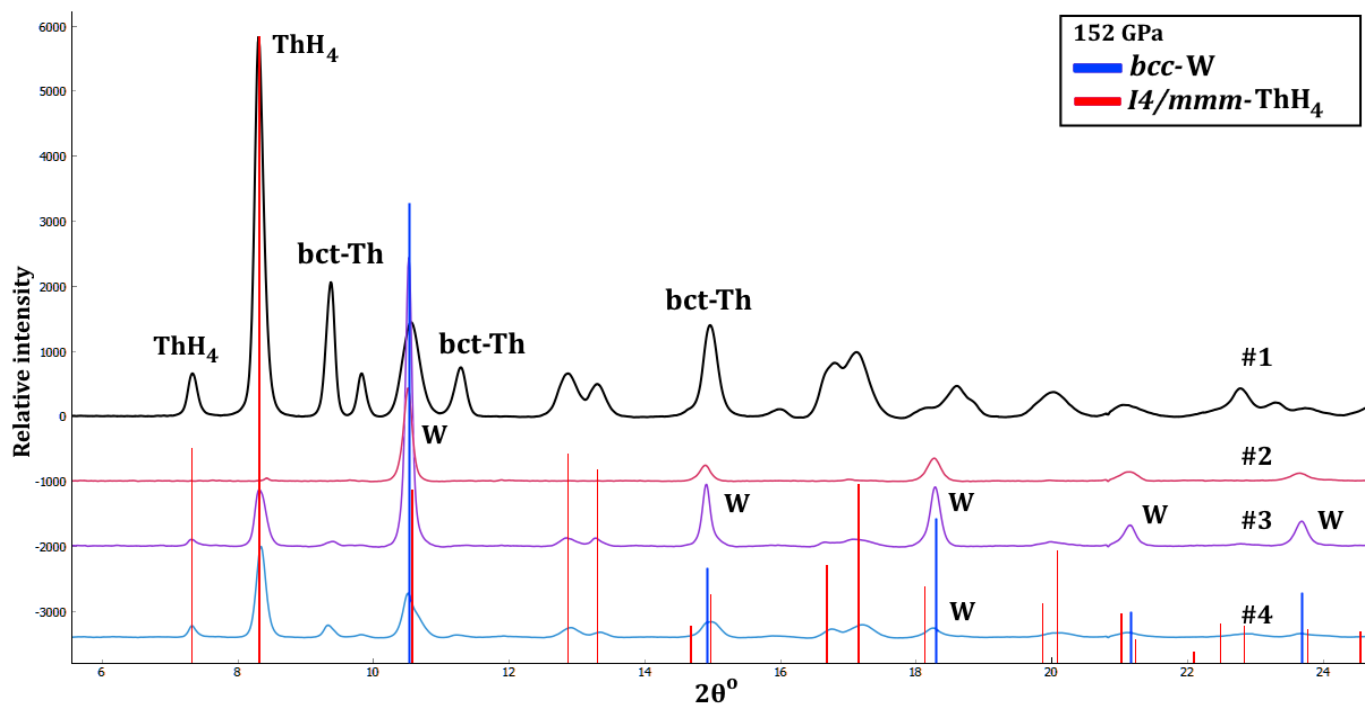

**Figure S3**. XRD patterns from different points ("cross") of M2 sample at 152 GPa (points #1-4). Only metallic *bcc*-W (gasket), *bct*-Th and *I*4/*mmm*-ThH$_4$ can be distinguished.

S5

# Synthesis of $Cmc2_1$-ThH$_6$

Our experiments suggest that the lower limit of the thermodynamic stability of ThH$_9$ is ~105-110 GPa, in perfect agreement with the DFT calculations (Figure S24). The calculations of the phonon spectra of $P6_3/mmc$-ThH$_9$ at 100-120 GPa (Figure S25) show that this phase is dynamically stable.

The measured X-ray diffraction patterns of M2 sample during the pressure release from 104 GPa to 86 GPa (Figure S4c) make it possible to conclude that ThH$_9$ participates in the following reaction:

$$\frac{2}{5}ThH_9 + \frac{3}{5}ThH_4 \rightarrow Cmc2_1 - ThH_6, \Delta H_0 = -0.075 \, eV/f.u.$$

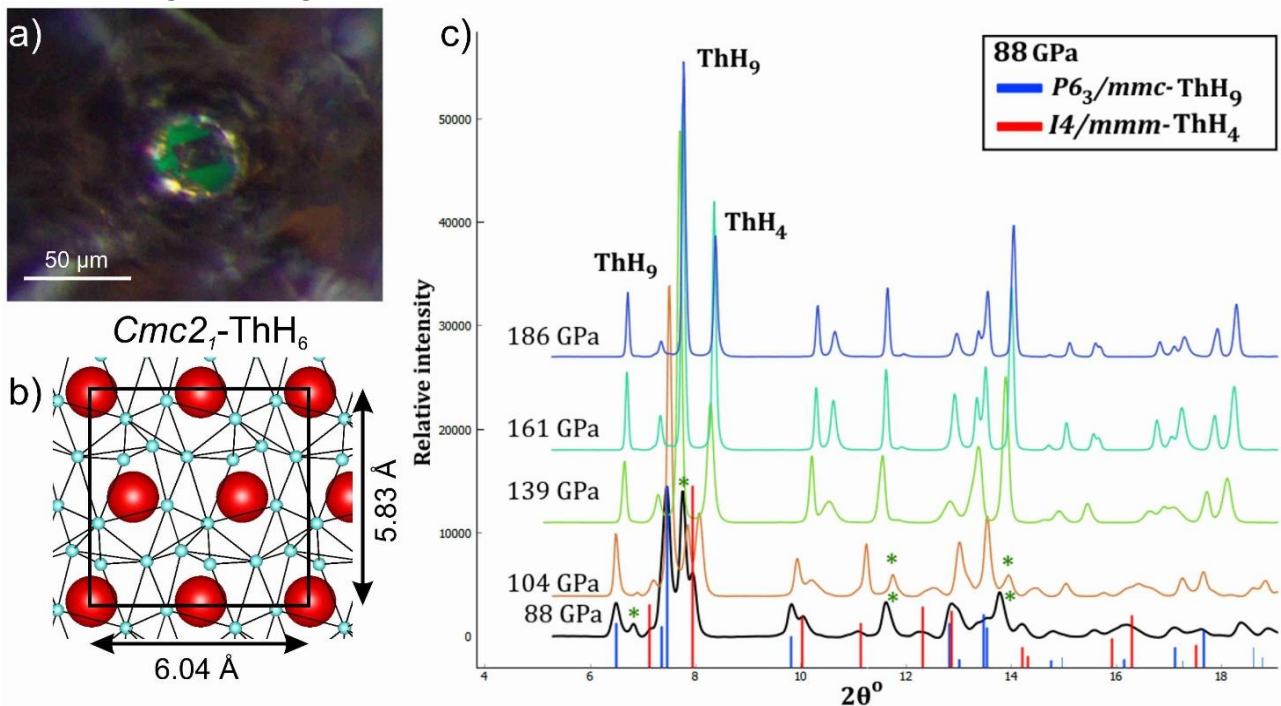

**Figure S4**. a) An optical image of M2 sample with cracking of diamonds and subsequent decrease in pressure to 30 GPa. Green space is a backlight from a 532-nm laser. b) Crystal structure of $Cmc2_1$-ThH$_6$ at 88 GPa. c) Combined XRD patterns of high-pressure phases in the Th-H system. During the pressure decrease from 104 to 88 GPa in the ThH$_4$+ThH$_9$ mixture, a new decomposition product forms (marked by green asterisks) with corresponding reflections (2θ°): 6.83, 7.77, 11.63, 12.68, 13.78 (at 88 GPa).

# Experimental lattice parameters of new phases

**Table S3**. Experimental lattice parameters and volume of *I4/mmm*-ThH$_4$.

| Pressure, GPa | $a$, Å | $c$, Å | $V$, Å$^3$ | $V_{DFT}$, Å$^3$ |
|---|---|---|---|---|
| 172 | 2.84(1) | 5.82(1) | 46.98(2) | 46.63 |
| 168 | 2.85(1) | 5.83(1) | 47.44(2) | 46.96 |
| 161 | 2.85(2) | 5.84(2) | 48.2(2) | 47.22 |
| 152 | 2.88(1) | 5.87(1) | 48.60(2) | 47.82 |
| 139 | 2.96(1) | 5.93(1) | 51.90(3) | 48.56 |
| 104 | 3.02(1) | 6.00 (1) | 54.71(1) | 52.78 |
| 86 | 3.06(1) | 6.07(1) | 56.98(2) | 54.96 |

**Table S4.** Experimental lattice parameters and volume of *P6$_3$/mmc*-ThH$_9$.

| Pressure, GPa | $a$, Å | $c$, Å | $V$, Å$^3$ | $V_{DFT}$, Å$^3$ |
|---|---|---|---|---|
| 172 | 3.678(1) | 5.475(3) | 64.14(3) | 63.61 |
| 168 | 3.687(1) | 5.487(3) | 64.62(3) | 64.25 |
| 161 | 3.706(2) | 5.522(3) | 65.70(8) | 64.59 |
| 152 | 3.713(1) | 5.541(3) | 66.20(3) | 65.32 |
| 139 | 3.809(1) | 5.683(5) | 71.39(5) | 66.42 |
| 104 | 3.810(1) | 5.820(1) | 73.17(2) | 70.78 |
| 86 | 3.878(1) | 5.875(3) | 76.51(3) | 72.24 |

**Table S5.** Experimental lattice parameters of $Fm\bar{3}m$-ThH$_{10}$.

| Pressure, GPa | a, Å | V, Å$^3$ | $V_{DFT}$, Å$^3$ |
|---|---|---|---|
| 183 | 5.09(3) | 131.84(5) | 130.68 |
| 171 | 5.11(1) | 133.52(5) | 131.96 |
| 151 | 5.13(1) | 135.11(7) | 135.00 |
| 143 | 5.15(1) | 136.99(2) | 136.88 |
| 132 | 5.18(1) | 139.24(5) | 138.88 |
| 120 | 5.22(1) | 141.82(7) | 141.04 |
| 101 | 5.26(1) | 145.92(1) | 145.85 |
| 85 | 5.29(1) | 148.00(1) | 150.18 |

**Table S6.** Experimental parameters of the third-order Birch-Murnaghan equation of state for newly synthesized phases.

| Phase | *bct*-Th | *I4/mmm*-ThH$_4$ | *Cmc2$_1$*-ThH$_6$ | *P6$_3$/mmc*-ThH$_9$ | $Fm\bar{3}m$-ThH$_{10}$ |
|---|---|---|---|---|---|
| $P_0$, GPa | 80 | 100 | 80 | 100 | 100 |
| $V_0$, Å$^3$ | 37.73 | 26.41 | 32.36 | 70.84 | 36.45 |
| $K_0$, GPa | 320.15 | 414.26 | 332.44 | 445.11 | 531.41 |
| $K_0'$ | 3.27 | 2.68 | 2.75 | 3.31 | 5.46 |

# Calculated and experimental EoS of Th-H phases

**Table S7.** Calculated EoS parameters of the third-order Birch–Murnaghan equation:

$$P = \frac{3}{4} K_0 \left(\frac{V}{V_0}\right)^{\frac{-5}{3}} \left(1 - \left(\frac{V}{V_0}\right)^{\frac{-2}{3}}\right) \cdot \left[\frac{3}{4}(K_0' - 4.37) \cdot \left(1 - \left(\frac{V}{V_0}\right)^{\frac{-2}{3}}\right) - 1\right]$$

| Phase | $Fm\bar{3}m$-Th | $I4/mmm$-Th[a] | $Fm\bar{3}m$-Au[b] | $Fm\bar{3}m$-W[c] | $Fm\bar{3}m$-Ta[d] |
|---|---|---|---|---|---|
| $V_0$ | 32.175 | 69.038 | 72.7030 | 31.930 | 36.239 |
| $K_0$ | 49.533 | 303.526 | 140.562 | 332.908 | 198.466 |
| $K'$ | 4.158 | 4.712 | 5.653 | 3.798 | 3.527 |

[a] $V_0 = 35.1278$ Å$^3$, $K_0 = 392.7$ GPa, $K_0' = 5.79$ [5].  [b] $V_0 = 67.847$ Å, $K_0 = 167.0$ GPa, $K_0' = 5.77$ [6].
[c] $V_0 = 31.70$ Å$^3$, $K_0 = 311.22$ GPa, $K_0' = 3.90$ [7].  [d] $V_0 = 34.278$ Å$^3$, $K_0 = 195.0$ GPa, $K_0' = 3.40$ [8].

**Table S8.** Experimental cell parameters for lower thorium hydrides and *bct*-Th

| Entry | Compound | Pressure, GPa | a, Å | b, Å | c, Å | V, Å$^3$ |
|---|---|---|---|---|---|---|
| 1 | *bct*-Th | 88 | 3.04(2) | | 4.39(4) | 40.68(4) |
| 2 | *bct*-Th | 98 | 2.901(5) | | 4.42(0) | 37.21(5) |
| 3 | *bct*-Th | 148 | 2.72(4) | | 4.37(3) | 32.45(4) |
| 4 | *bct*-Th | 152 | 2.68(7) | | 4.37(7) | 31.61(7) |
| 5 | *bct*-Th | 182 | 2.65(9) | | 3.92(1) | 27.74(9) |
| 6 | $P321$-ThH$_4$ | 88 | 5.50(1) | | 3.29(1) | 86.18(3) |
| 7 | $Cmc2_1$-ThH$_6$ | 86 | 3.69(1) | 6.02(1) | 5.72(1) | 126.93(2) |
| 8 | $Cmc2_1$-ThH$_6$ | 104 | 3.64(2) | 5.98(2) | 5.72(2) | 124.70(9) |

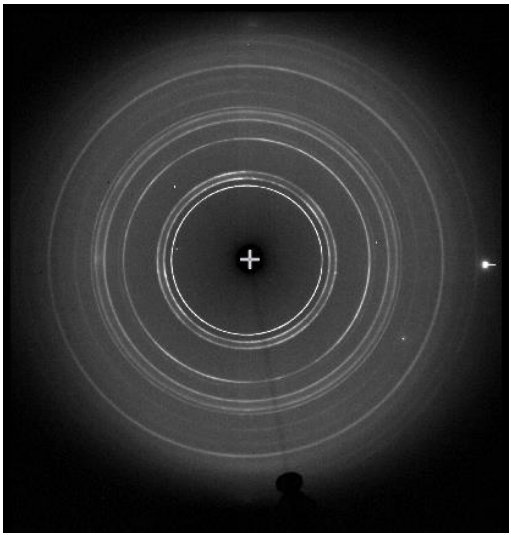 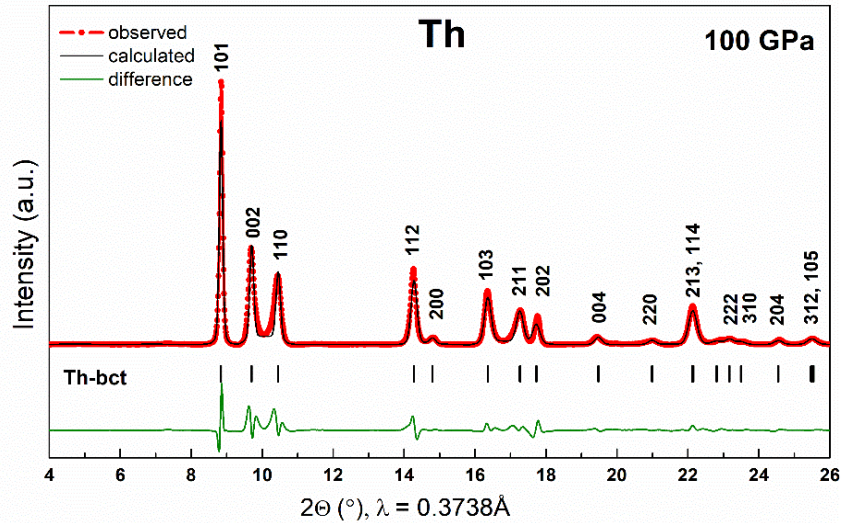

**Figure S5.** XRD image and pattern of M2 cell with a Th sample at ~100 GPa before the experiment. Only the reflections of metallic Th were detected.

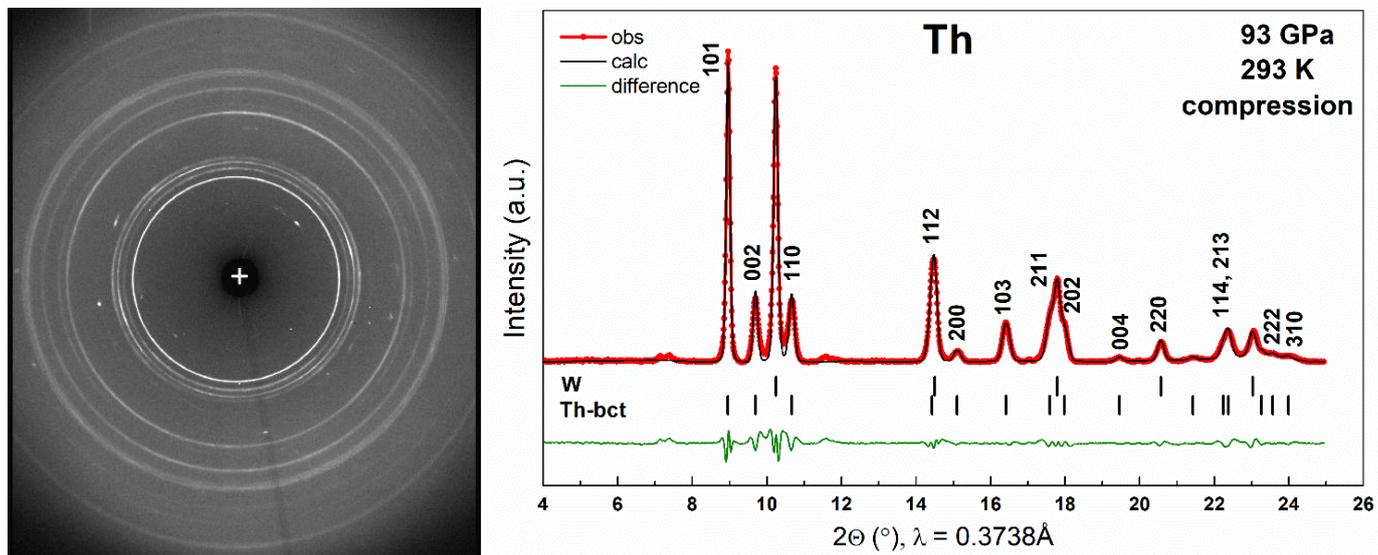

**Figure S6**. XRD image and pattern of M3 cell with a Th sample before the experiment. Only the reflections of metallic Th and W were detected.

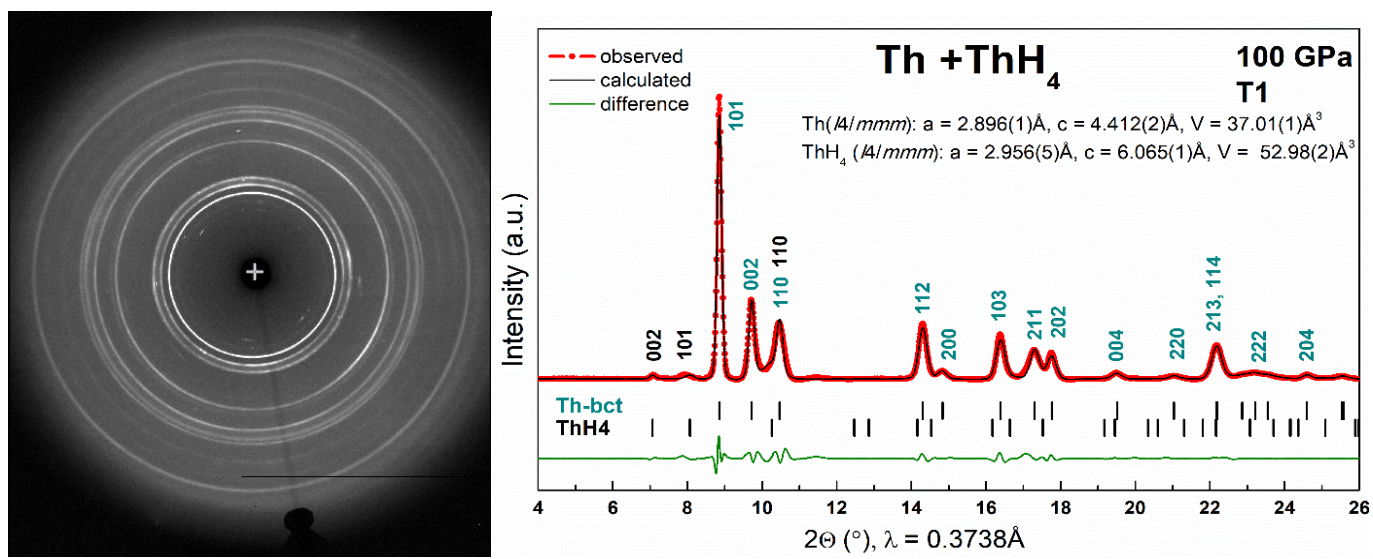

**Figure S7**. XRD image and pattern of M2 cell with a Th sample after the first heating (Run 1, 88 GPa). Only the reflections of metallic Th and $I4/mmm$-ThH$_4$ were detected.

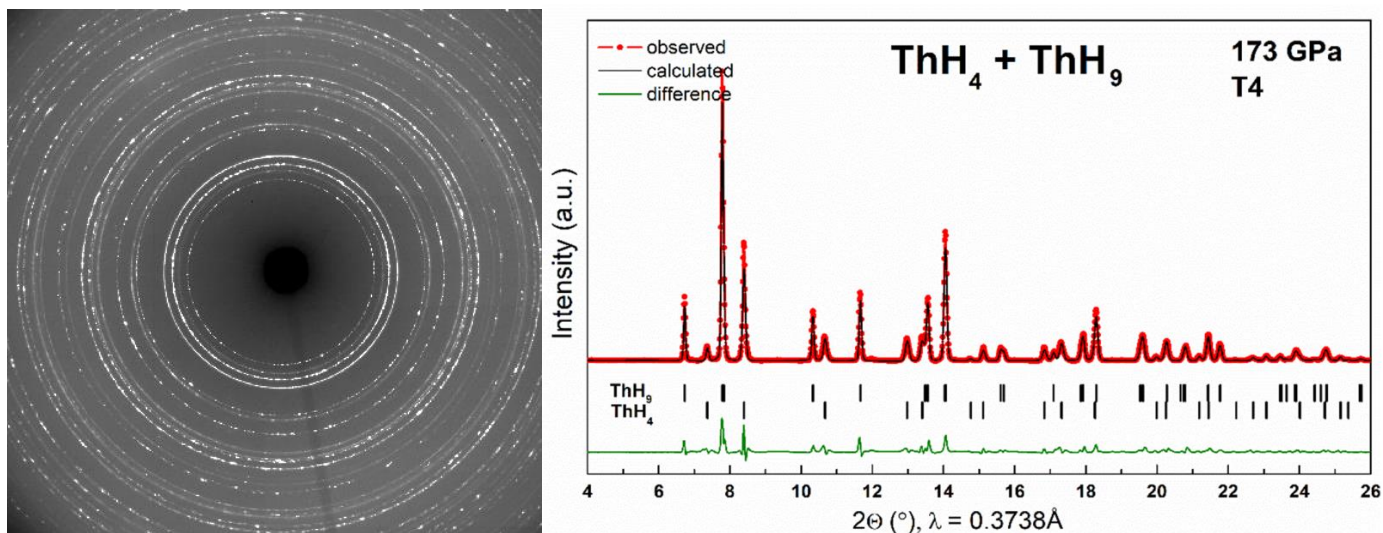

**Figure S8**. XRD image and pattern of M2 cell with a Th sample after the fourth heating (Run 3, 173 GPa). Only the reflections of $I4/mmm$-ThH$_4$ and $P6_3/mmc$-ThH$_9$ (dominant phase) were detected.

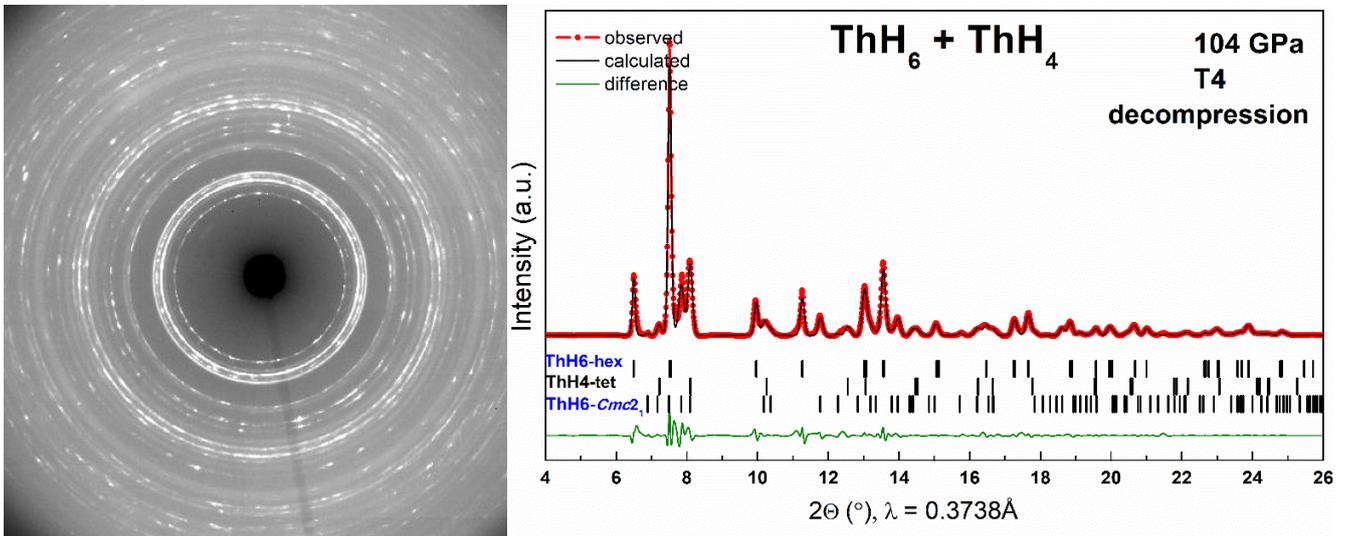

**Figure S9**. XRD image and pattern of M2 cell with a Th sample after the fourth heating and decompression (Run 4, 104 GPa). The new additional reflections match $Cmc2_1$-ThH$_6$.

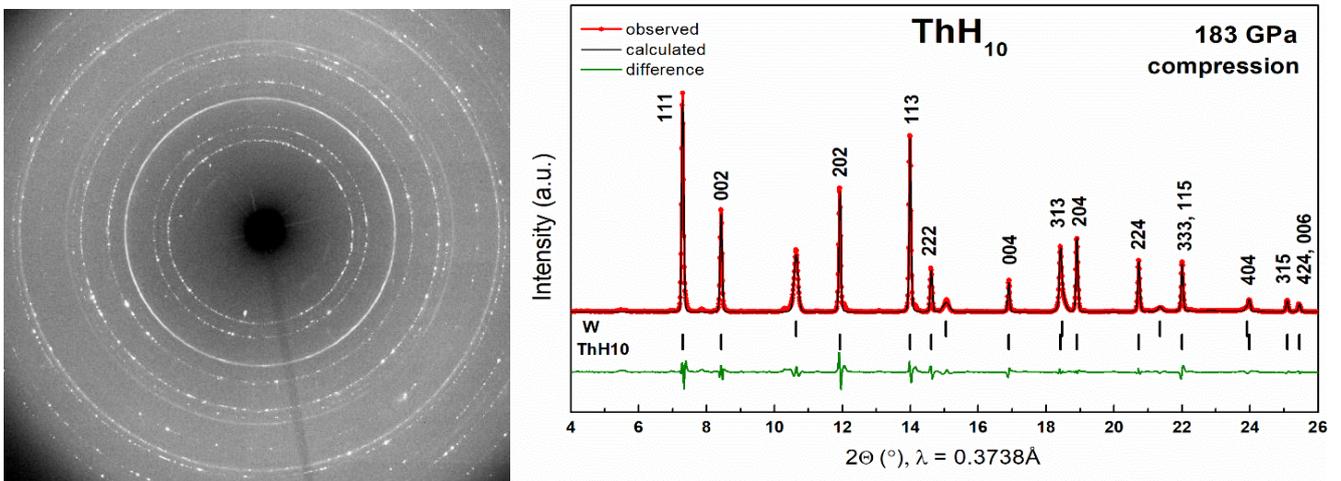

**Figure S10**. XRD image and pattern of M3 cell with a Th sample after the first heating (Run 5, 183 GPa). Only the reflections of $Fm\bar{3}m$-ThH$_{10}$ and W gasket were detected.

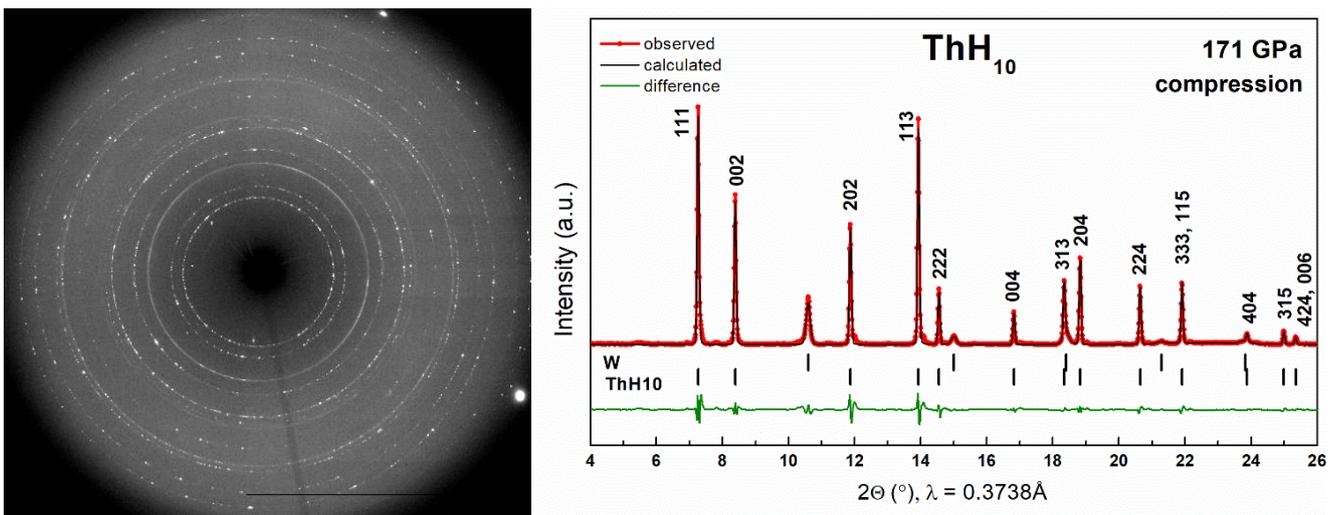

**Figure S11**. XRD image and pattern of M3 cell with a Th sample after the first heating (Run 5, 171 GPa). Only the reflections of $Fm\bar{3}m$-ThH$_{10}$ and W gasket were detected.

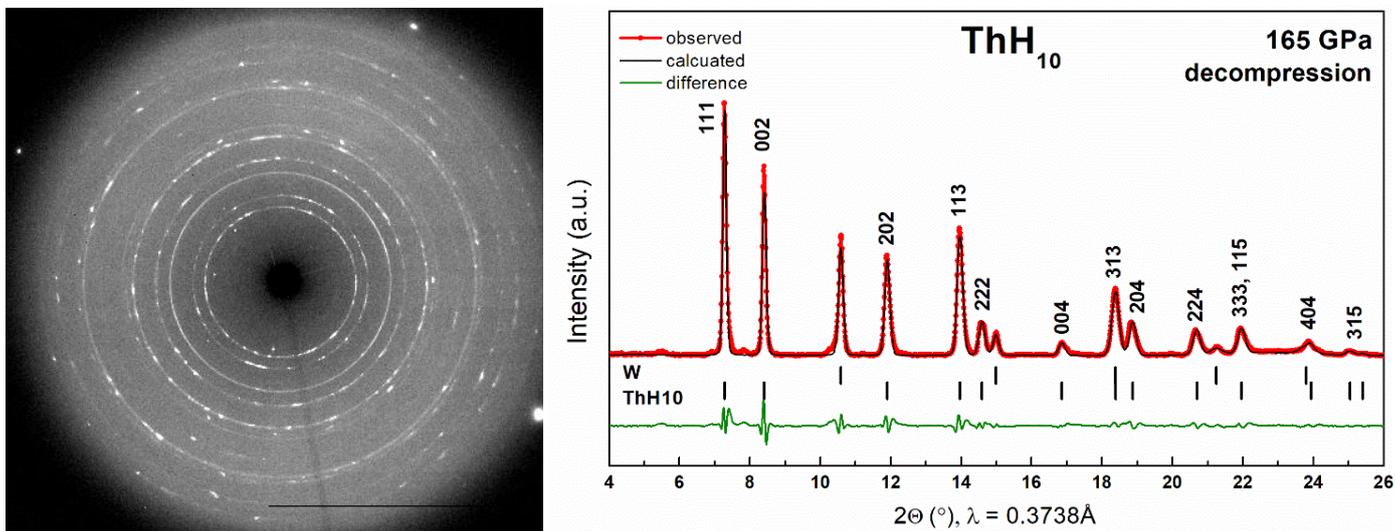

**Figure S12**. XRD image and pattern of M3 cell with a Th sample after the first heating (Run 5, 165 GPa). Only the reflections of $Fm\bar{3}m$-ThH$_{10}$ and W gasket were detected.

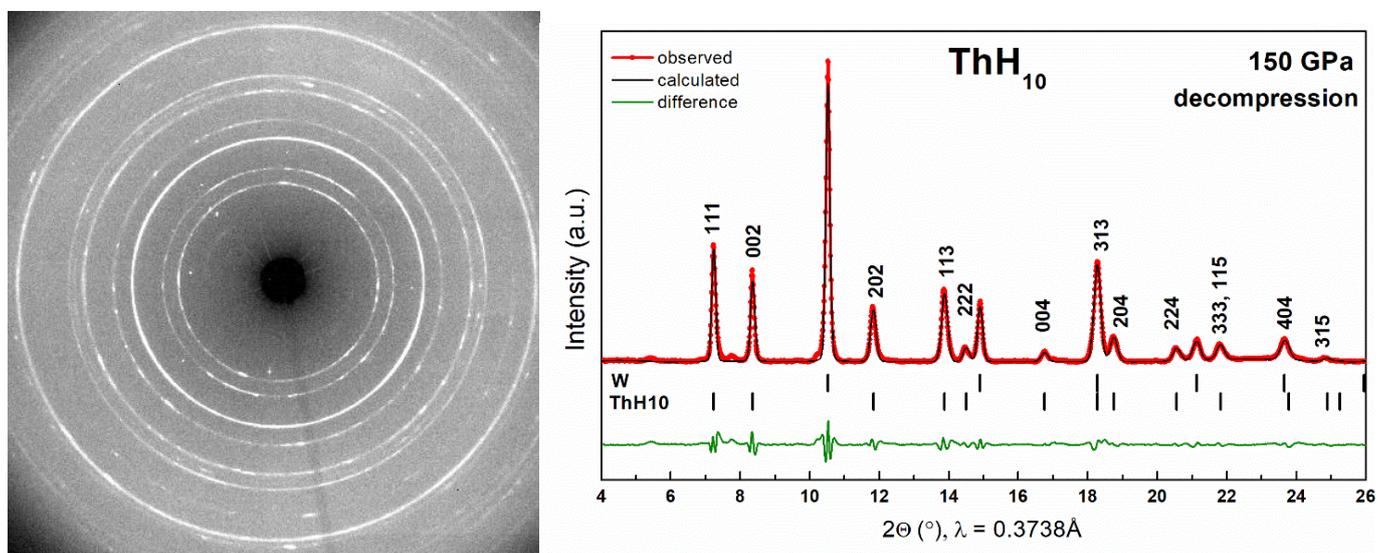

**Figure S13**. XRD image and pattern of M3 cell with a Th sample after the first heating (Run 5, 150 GPa). Only the reflections of $Fm\bar{3}m$-ThH$_{10}$ and W gasket were detected.

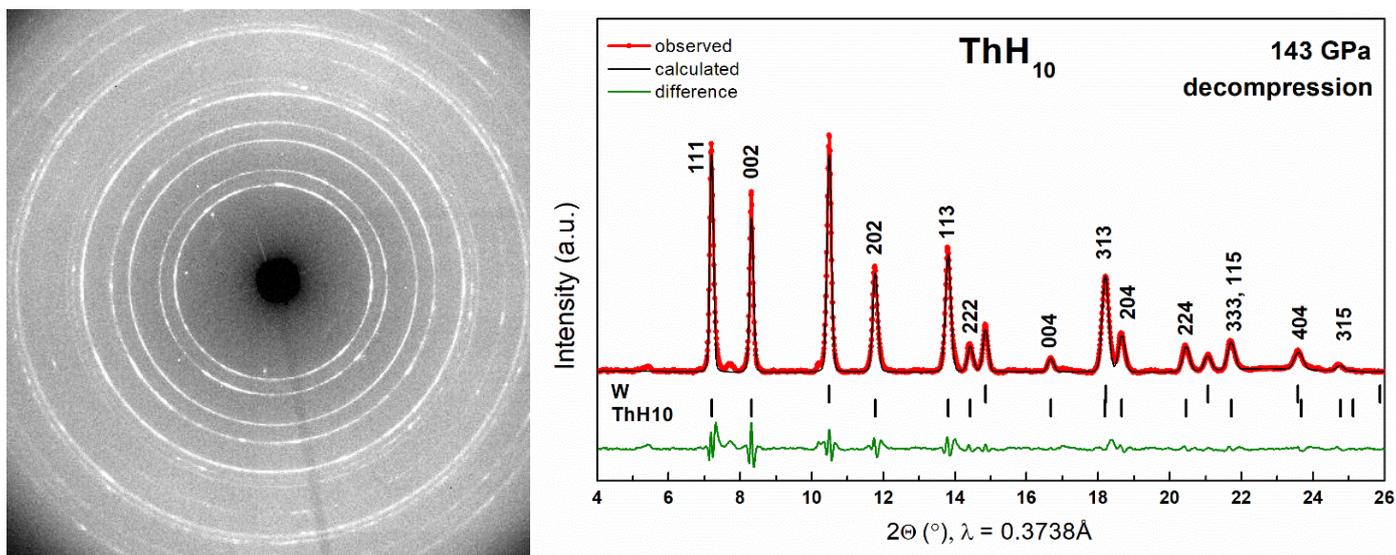

**Figure S14**. XRD image and pattern of M3 cell with a Th sample after the first heating (Run 5, 143 GPa). Only the reflections of $Fm\bar{3}m$-ThH$_{10}$ and W gasket were detected.

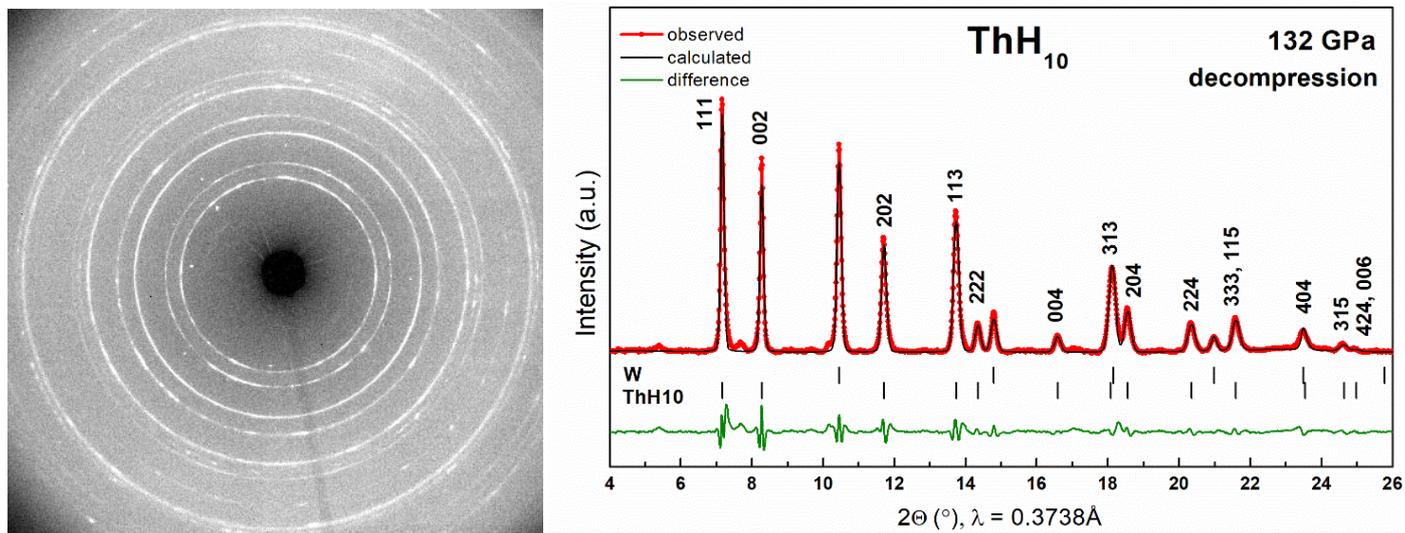

**Figure S15.** XRD image and pattern of M3 cell with a Th sample after the first heating (Run 5, 132 GPa). Only the reflections of $Fm\bar{3}m$-ThH$_{10}$ and W gasket were detected.

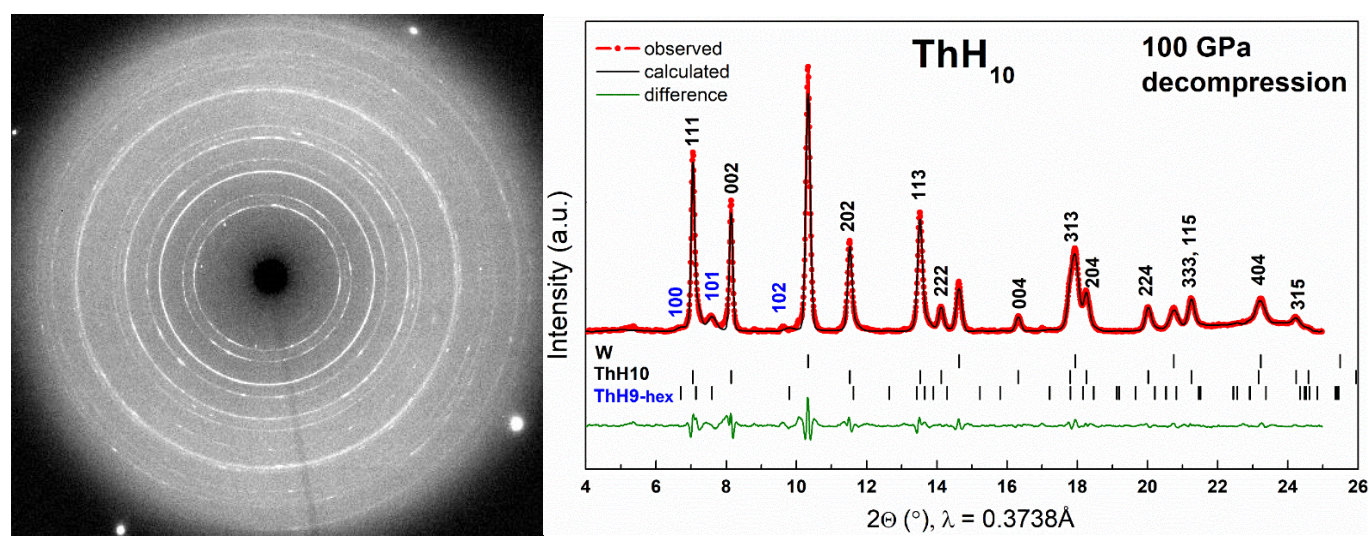

**Figure S16.** XRD image and pattern of M3 cell with a Th sample after the first heating (Run 5, 100 GPa). The reflections of $Fm\bar{3}m$-ThH$_{10}$, $P6_3/mmc$-ThH$_9$ and W gasket were detected.

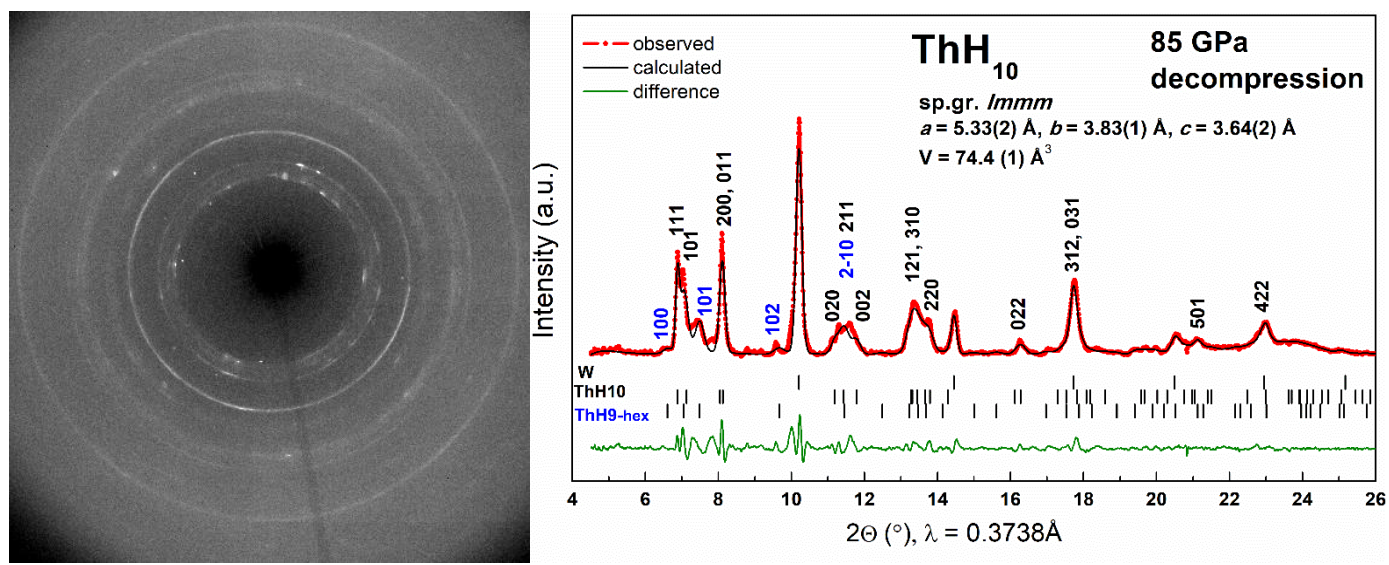

**Figure S17.** XRD image and pattern of M3 cell with a Th sample after the decomposition of $Fm\bar{3}m$-ThH$_{10}$ (Run 5, 85 GPa). The reflections of $Immm$-ThH$_{10}$, $P6_3/mmc$-ThH$_9$ and W gasket were detected.

# Rietveld refinements for ThH₄, ThH₉ and ThH₁₀

The refinement of obtained samples was made using FullProf software [9] having an extended set of specialized functions. During the refinement, the modified pseudo-Voigt function was used with several additional variable parameters, in particular the parameters of anisotropic peak broadening, which is strongly manifested in X-rays with a short wavelength. The refinement was made of the crystal structure parameters of ThH₄+ThH₉ (lattice parameters, $B_{iso}$ for Th atoms), microstructural parameters (anisotropic strain broadening), and texture in the March–Dollase model (Figure S18), which leads to satisfactory R-factor values.

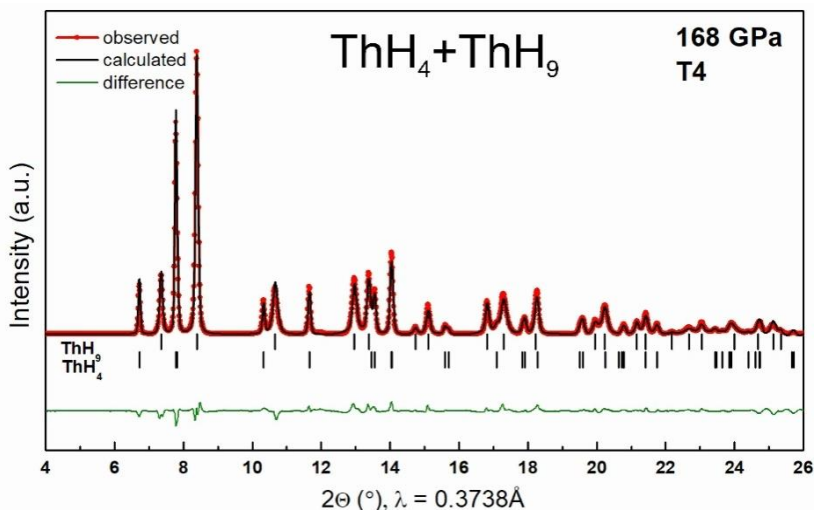

**Figure S18.** The Rietveld refinement of $P6_3/mmc$-ThH₉ and ThH₄ at 168 GPa. The experimental data are shown in red, a fit approximation and residues are denoted by black and green lines, respectively. The R-factors (not corrected for background) for ThH₄+ThH₉ are: $R_p = 6.05\%$, $R_{wp} = 8.04\%$, $R_{exp} = 4.85\%$, $Chi_2 = 2.75\%$. The conventional Rietveld R-factors for ThH₄+ThH₉ are: $R_p = 0.9\%$, $R_{wp} = 12.8\%$, $R_{exp} = 7.68\%$, $Chi_2 = 2.75\%$.

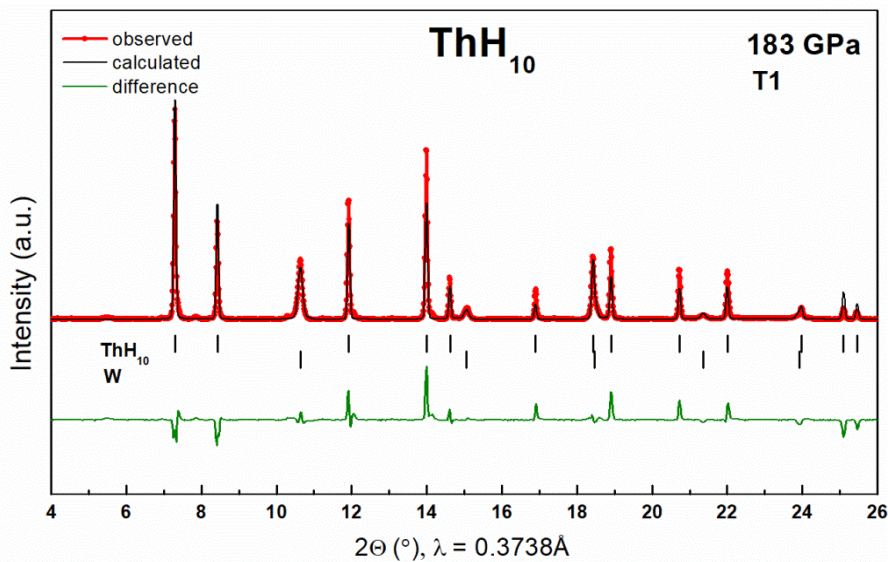

**Figure S19.** The Rietveld refinement of $Fm\bar{3}m$-ThH₁₀ and bcc-W at 183 GPa. The R-factors (not corrected for background) for ThH₁₀ are: $R_p = 8.64\%$, $R_{wp} = 13.9\%$, $R_{exp} = 5.59\%$, $Chi_2 = 6.22\%$. The conventional Rietveld R-factors for ThH₁₀ are: $R_p = 29.6\%$, $R_{wp} = 30.6\%$, $R_{exp} = 12.28\%$, $Chi_2 = 6.22\%$.

The intensities of the ThH₁₀ reflections are very distorted, and the Rietveld refinement is not satisfactory. Strong mechanical deformations and the pronounced texture of a sample at such high pressures inevitably lead to distortions of the reflex intensities. Another possible source of significant distortions is the X-ray absorption by DAC, the correction for which is extremely difficult to introduce in this case. In addition, it is impossible to clearly determine the positions of hydrogen atoms on the background of heavy Th atoms occupying fixed positions during refinement using X-ray diffraction patterns in the studied structures of hydrides. Even with single-crystal X-ray data, it is not always possible to determine the exact positions of hydrogen atoms, in which case they are presumed from simulation data.

# Stability of thorium hydrides

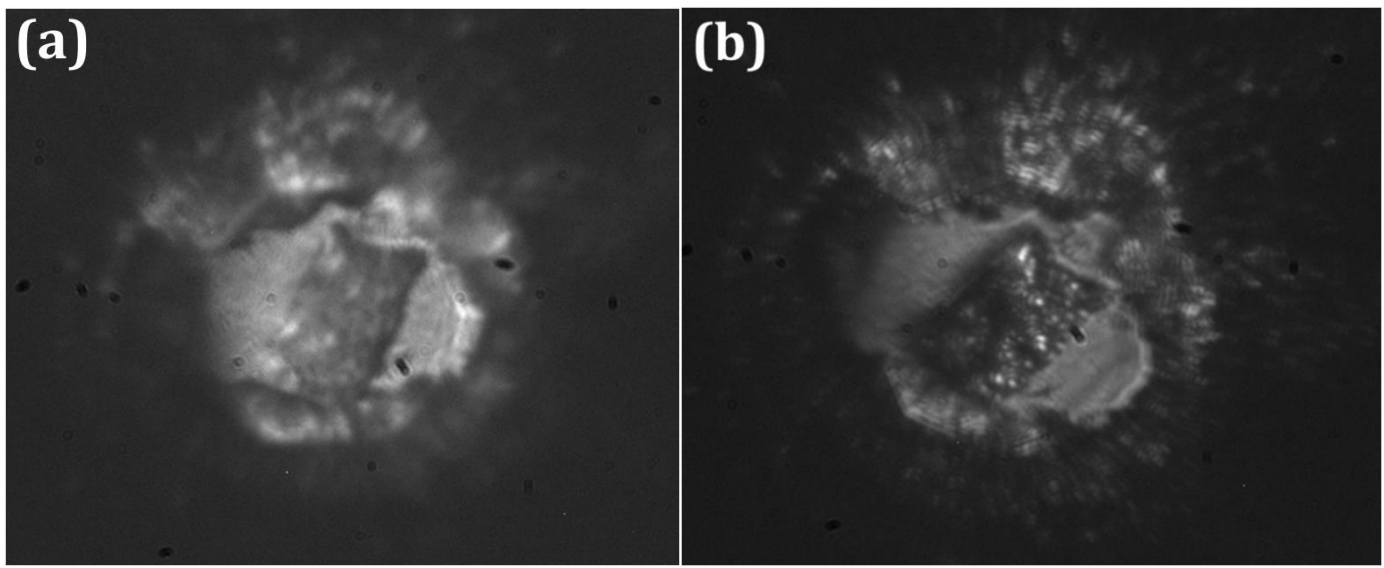

**Figure S20**. Sample M2: (a) before the laser heating; (b) after the laser heating of the left corner.

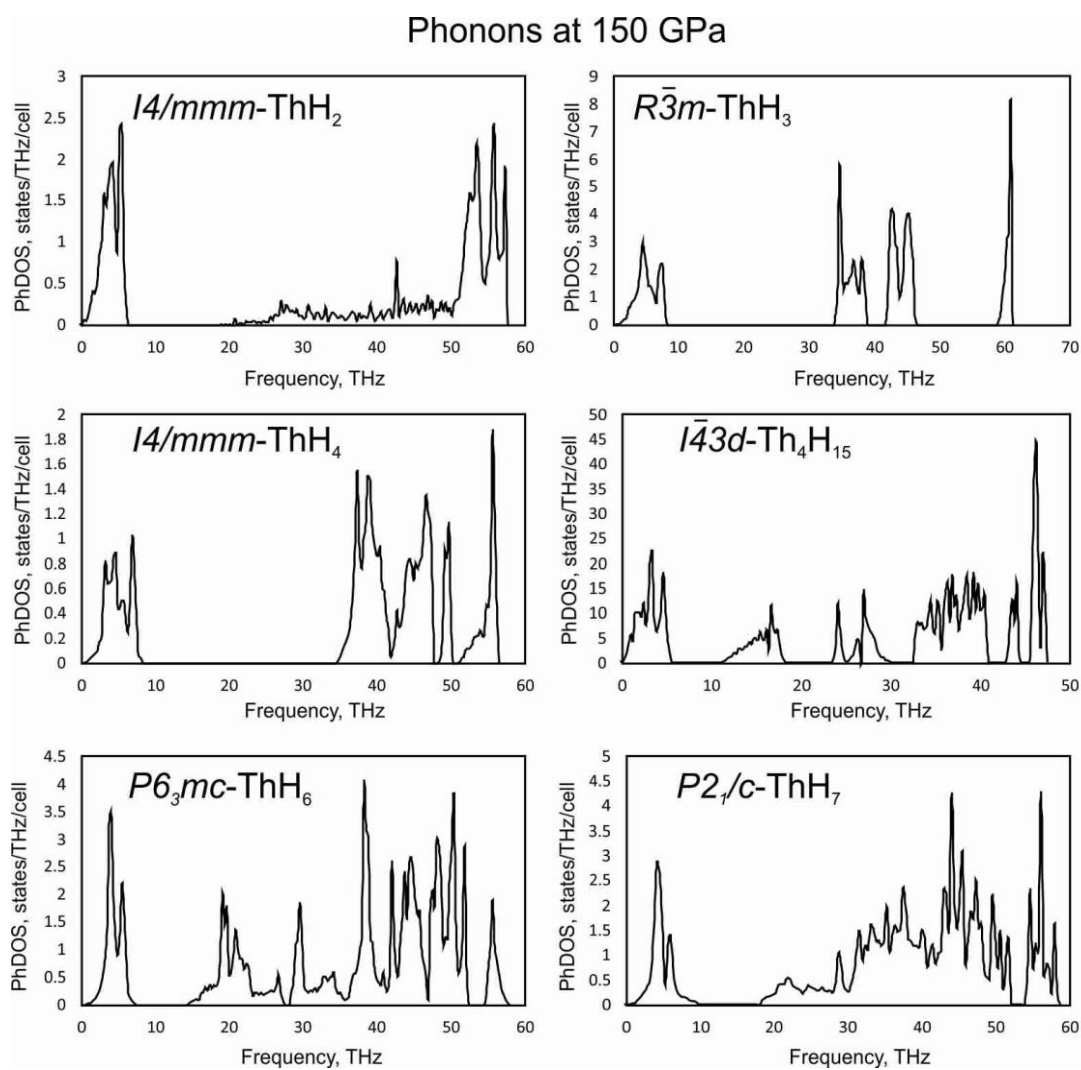

**Figure S21**. Calculated phonon DOS of lower thorium hydrides at 150 GPa demonstrating dynamical stability of all phases. These spectra were used for the ZPE and entropy calculations.

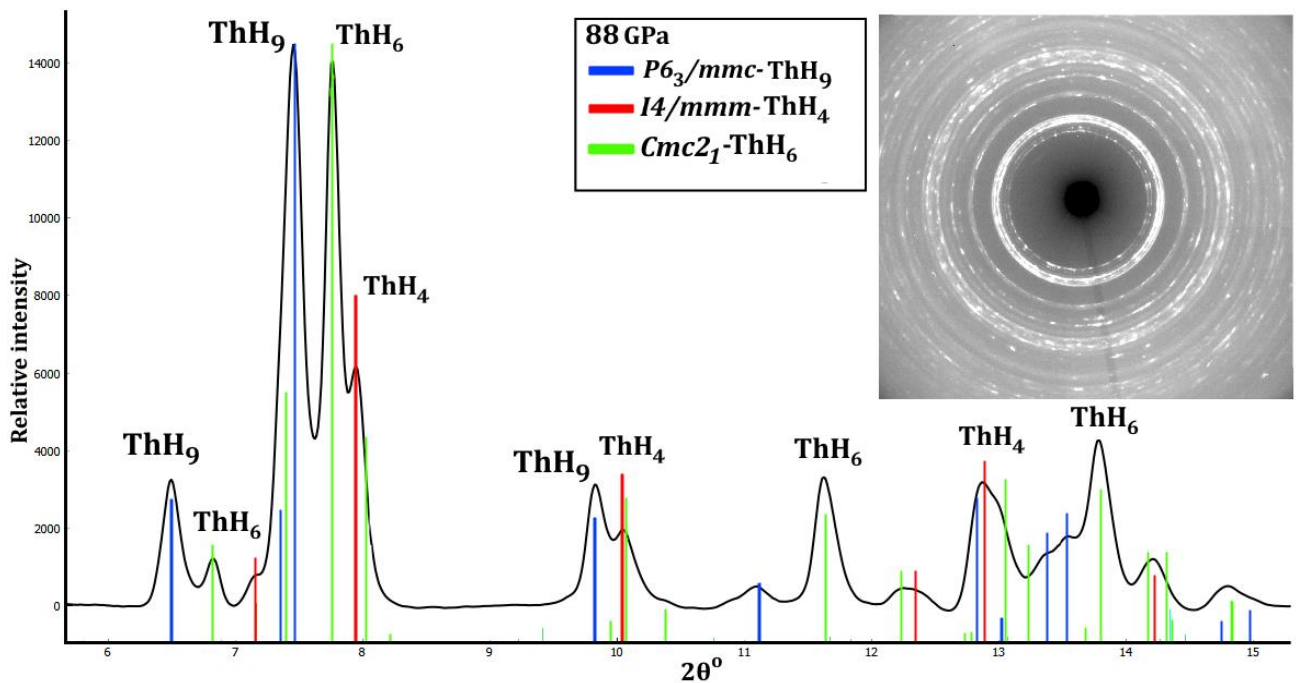

**Figure S22**. Combined XRD patterns of high-pressure phases in the Th-H system at 88 GPa (M2 sample, Run 4). Coexistence of thorium hydrides from $P6_3/mmc$-ThH$_9$ to $Cmc2_1$-ThH$_6$ is observed.

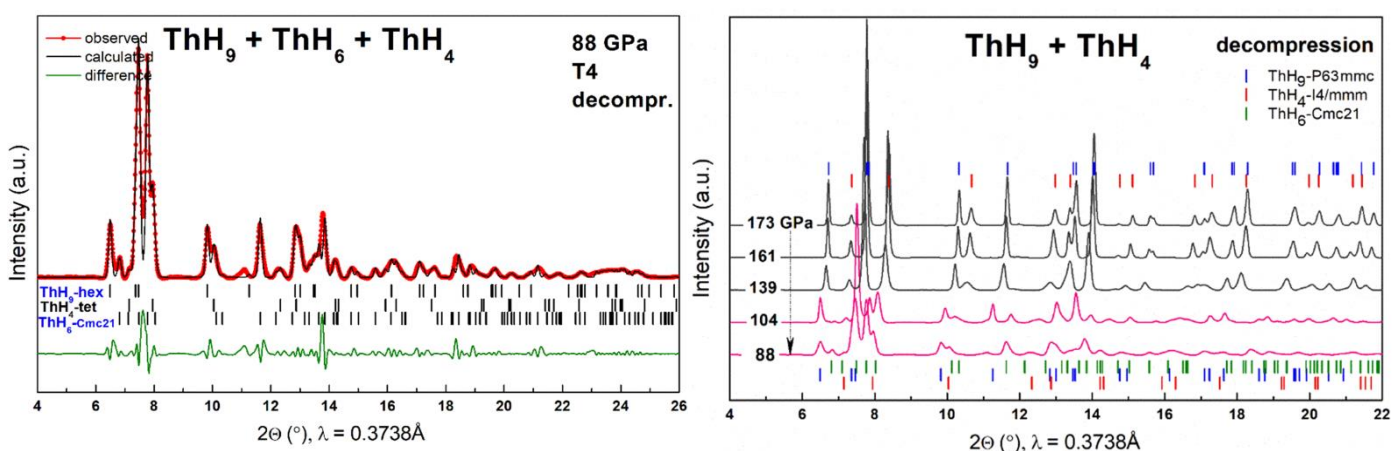

**Figure S23**. (left) The Le Bail refinement of $P6_3/mmc$-ThH$_9$, $I4/mmm$-ThH$_4$, and $Cmc2_1$-ThH$_6$ at 86 GPa (M2 sample). The experimental data are shown in red, a fit approximation and residue are denoted by black and green lines, respectively. (right) The evolution of XRD patterns of ThH$_9$, ThH$_6$, and ThH$_4$ between 173 and 88 GPa.

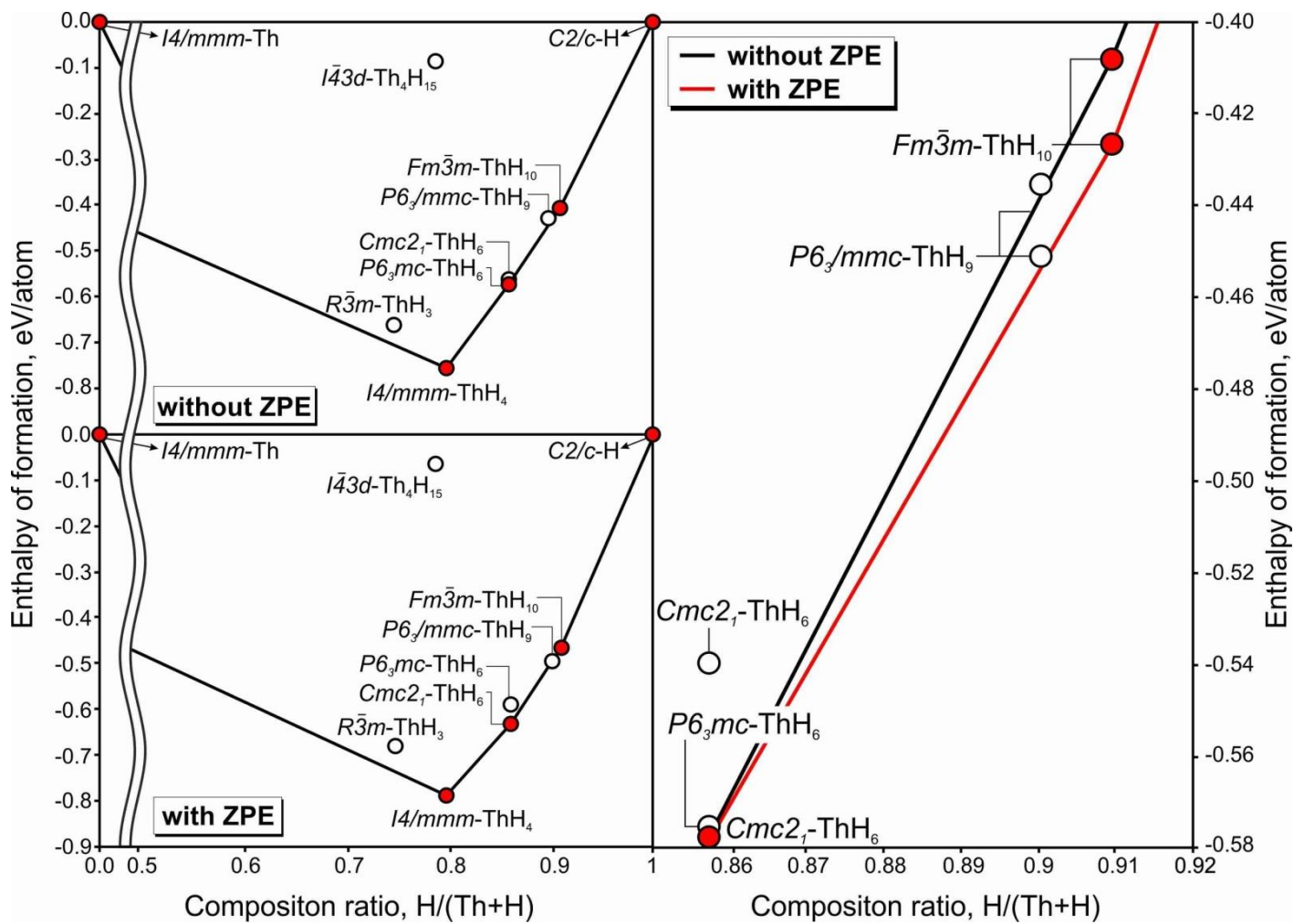

**Figure S24**. A convex hull diagram of the Th-H system at 100 GPa with (red line) and without (black line) the ZPE correction. At this pressure, ThH$_9$ no longer a stable compound (above the convex hull) even with the ZPE correction because of the emergence of ThH$_6$ on the convex hull.

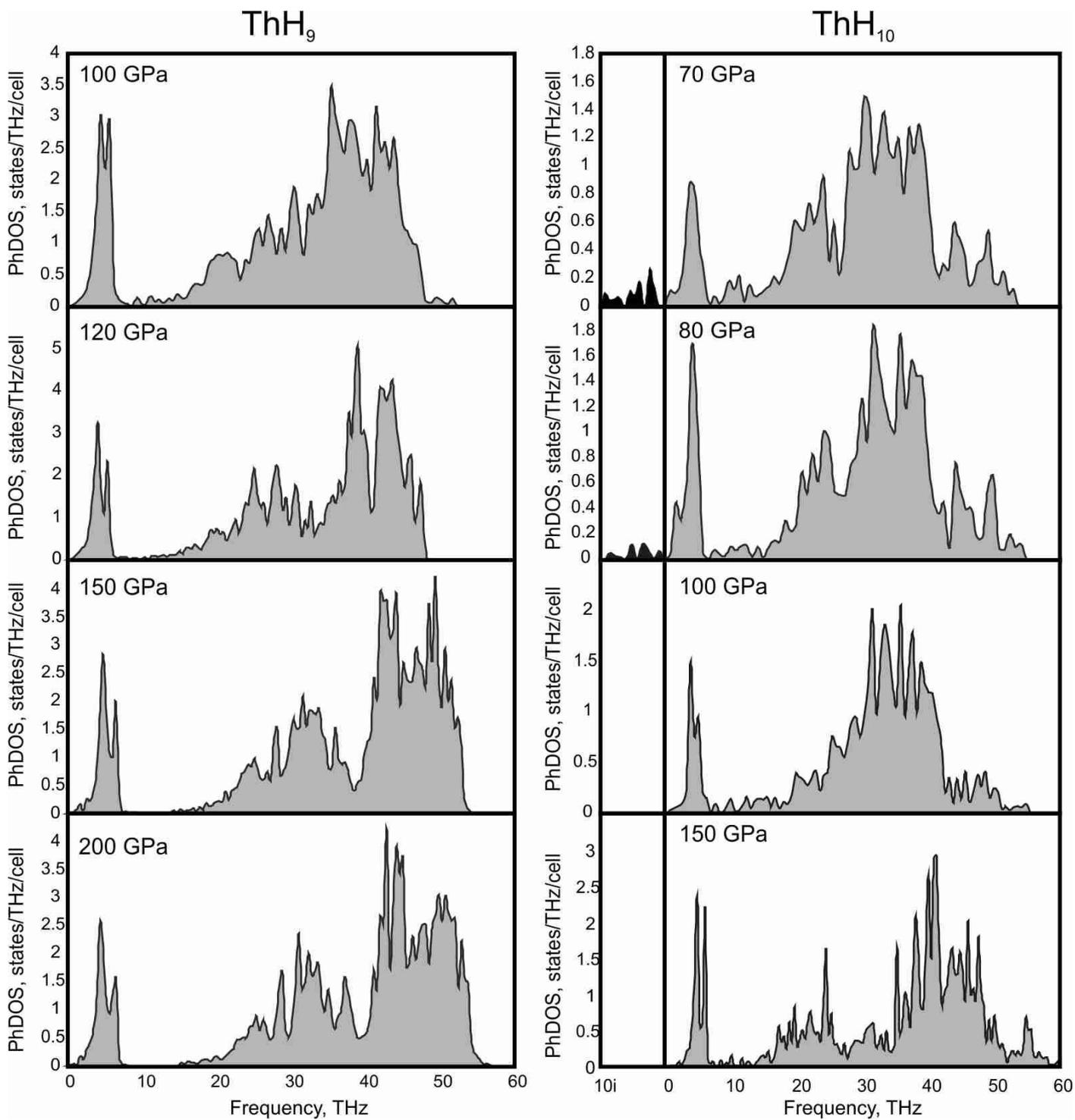

**Figure S25**. The calculated phonon DOS of *P6₃/mmc*-ThH₉ and $Fm\bar{3}m$-ThH₁₀ demonstrating dynamic stability of ThH₉ in the range of 100-200 GPa and instability of ThH₁₀ at and below 80 GPa.

**Decompression**

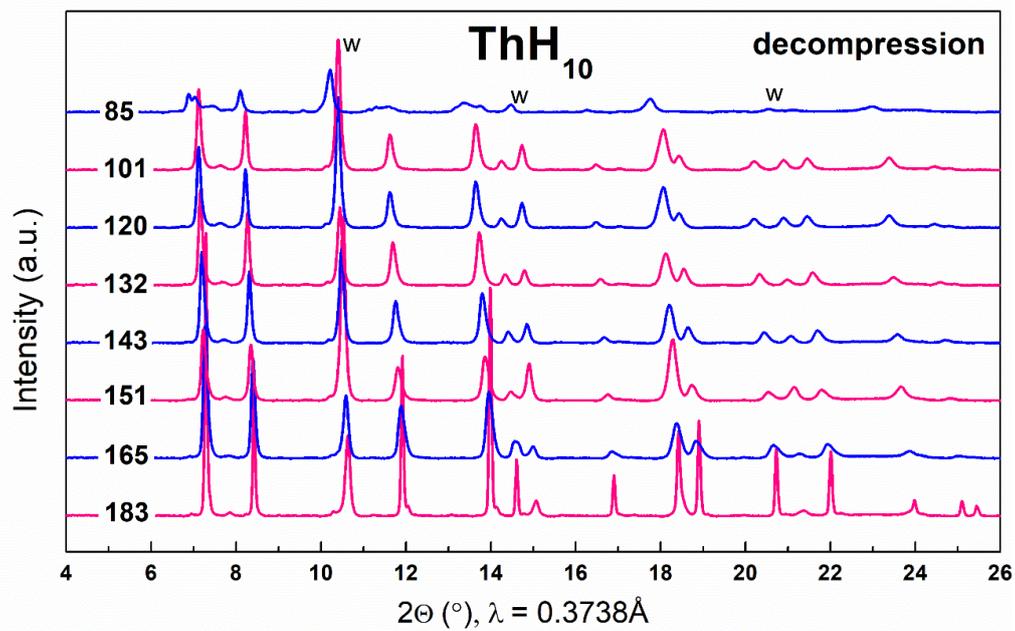

**Figure S26**. Experimental XRD patterns of $Fm\bar{3}m$-ThH$_{10}$ in the pressure range of 183-85 GPa.

# Other experimental data

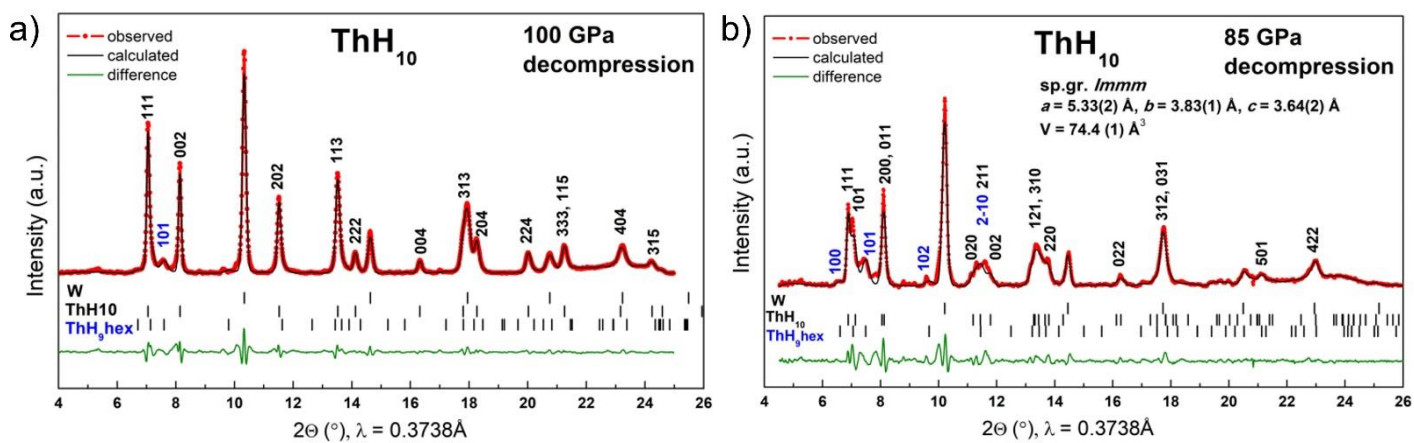

**Figure S27.** The Le Bail refinement of $Fm\bar{3}m$-ThH$_{10}$ and $bcc$-W at a) 100 GPa and b) 85 GPa.

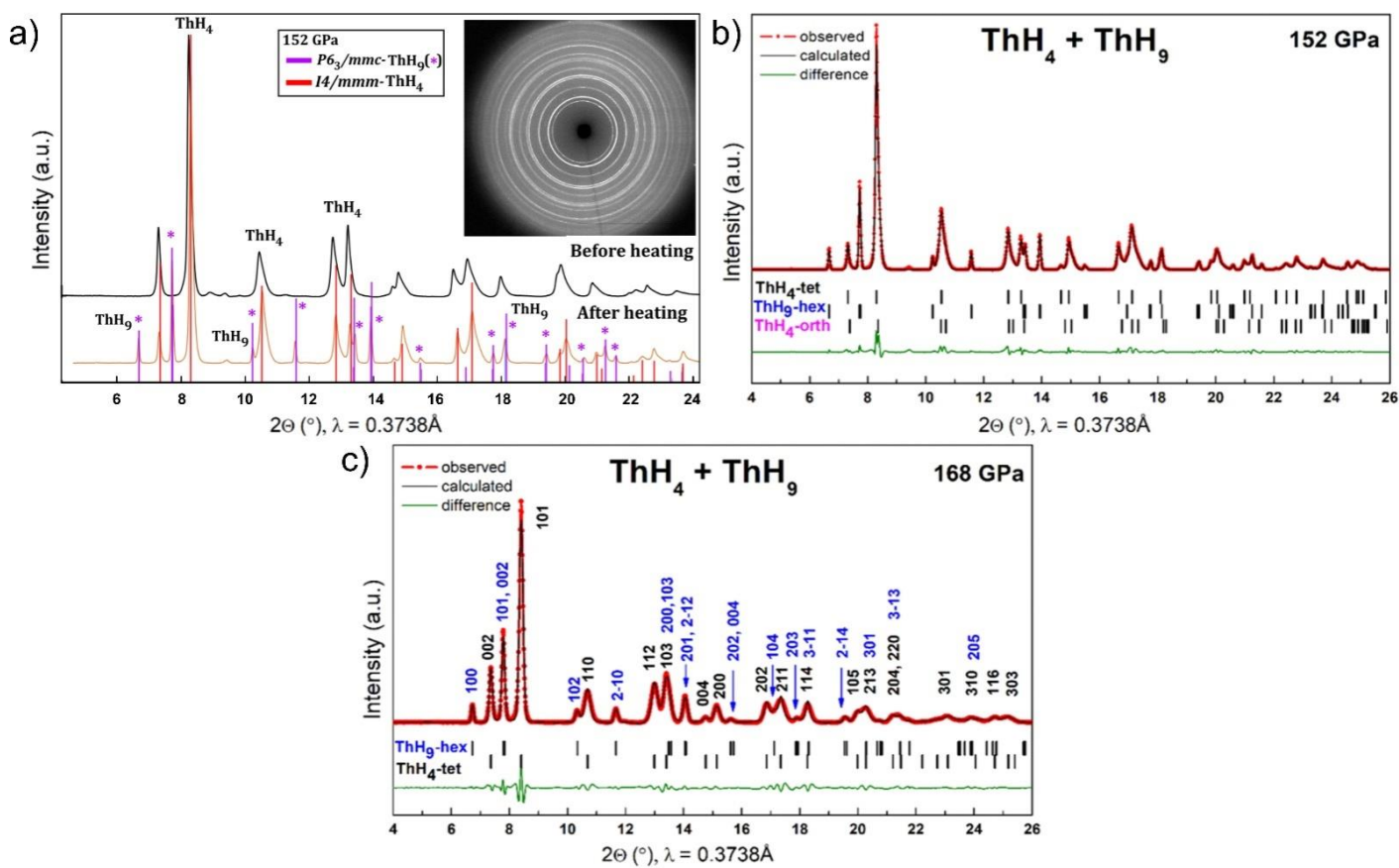

**Figure S28.** a) XRD patterns of M2 sample at 152 GPa before and after heating. The formation of ThH$_9$ is observed. The new peaks are highlighted by asterisks (*). The experimental XRD image obtained with the incident X-ray wavelength of 0.3738 Å is shown in an inset.

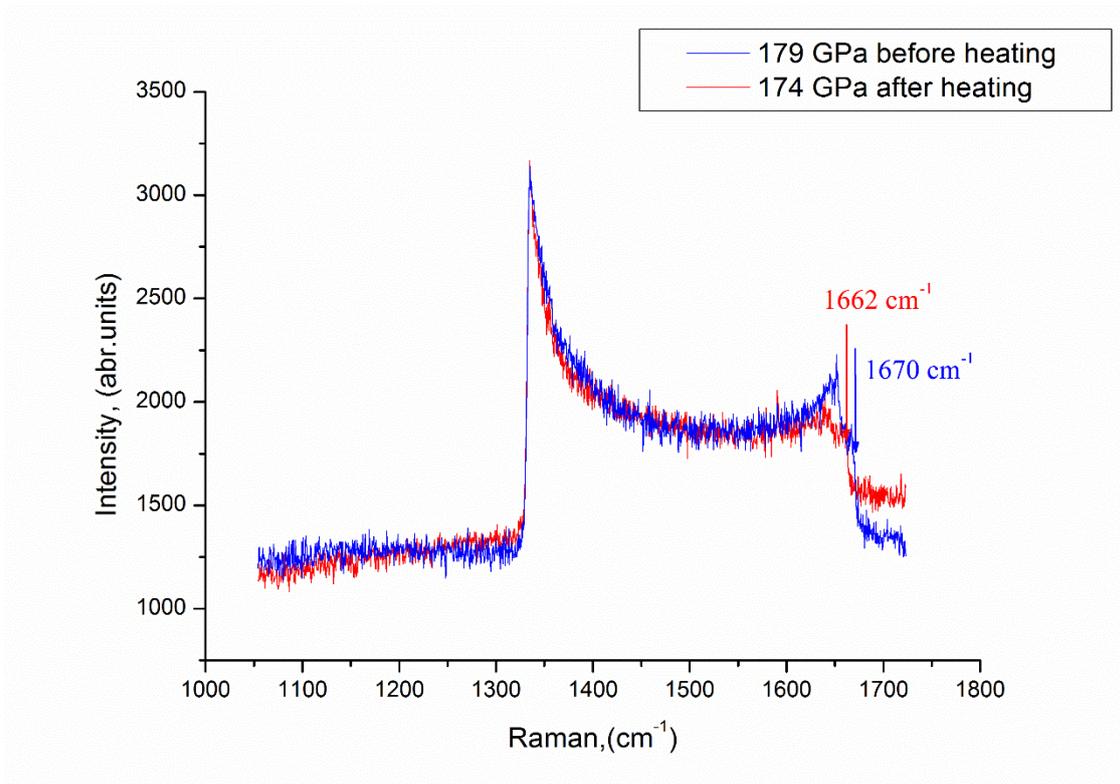

**Figure S29.** Raman spectra of the DAC prepared for the resistance measurements of ThH$_{10}$.

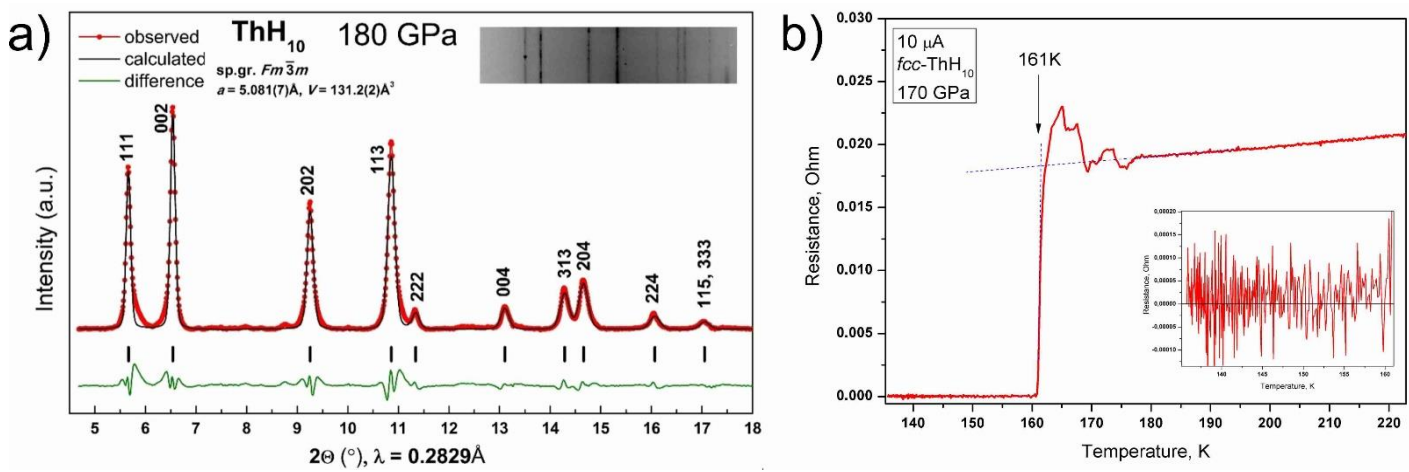

**Figure S30.** a) The Le Bail refinement of $Fm\bar{3}m$-ThH$_{10}$ at 180 GPa. The experimental data are shown in red, a fit approximation and residue are denoted by black and green lines, respectively. b) The temperature dependence of the resistance of ThH$_{10}$.

# Equations for calculating $T_C$ and related parameters

The critical temperature of superconducting transition was calculated using the Matsubara-type linearized Eliashberg equations: [10]

$$\hbar\omega_j = \pi(2j+1) \cdot k_B T, j = 0, \pm 1, \pm 2, \ldots \tag{S1}$$

$$\lambda(\omega_i - \omega_j) = 2\int_0^\infty \frac{\omega \cdot \alpha^2 F(\omega)}{\omega^2 + (\omega_i - \omega_j)^{\wedge}2} d\omega \tag{S2}$$

$$\Delta(\omega = \omega_i, T) = \Delta_i(T)$$
$$= \pi k_B T \sum_j \frac{[\lambda(\omega_i - \omega_j) - \mu^*]}{\rho + |\hbar\omega_j + \pi k_B T \sum_k (sign\,\omega_k) \cdot \lambda(\omega_i - \omega_j)|} \cdot \Delta_j(T) \tag{S3}$$

where $T$ is the temperature in kelvins, $\mu^*$ is the Coloumb pseudopotential, $\omega$ is the frequency in Hz, $\rho(T)$ is a pair-breaking parameter, the function $\lambda(\omega_i - \omega_j)$ relates to an effective electron-electron interaction *via* the exchange of phonons. [11] The transition temperature can be found as the solution of the equation $\rho(T_C) = 0$, where $\rho(T)$ is defined as $max(\rho)$, provided that $\Delta(\omega)$ is not a zero function of $\omega$ at a fixed temperature.

These equations can be rewritten in a matrix form as [12]

$$\rho(T)\psi_m = \sum_{n=0}^{N} K_{mn}\psi_n \Leftrightarrow \rho(T)\begin{pmatrix}\psi_1 \\ \ldots \\ \psi_N\end{pmatrix} = \begin{pmatrix}K_{11} & \ldots & K_{1N} \\ \ldots & K_{ii} & \ldots \\ K_{N1} & \ldots & K_{NN}\end{pmatrix} \times \begin{pmatrix}\psi_1 \\ \ldots \\ \psi_N\end{pmatrix}, \tag{S4}$$

where $\psi_n$ relates to $\Delta(\omega, T)$, and

$$K_{mn} = F(m-n) + F(m+n+1) - 2\mu* - \delta_{mn}\left[2m + 1 + F(0) + 2\sum_{l=1}^{m} F(l)\right] \tag{S5}$$

$$F(x) = F(x,T) = 2\int_0^{\omega max} \frac{\alpha^2 F(\omega)}{(\hbar\omega)^2 + (2\pi \cdot k_B T \cdot x)^2} \hbar\omega d\omega, \tag{S6}$$

where $\delta_{nn} = 1$ and $\delta_{nm} = 0$ ($n \neq m$) is a unit matrix. Now we can replace the criterion of $\rho(T_C) = 0$ by the vanishing of the maximum eigenvalue of the matrix $K_{nm}$: $\{\rho = max\_eigenvalue(K_{nm}) = f(T), f(T_C) = 0\}$.

Results of numerical solution of isotropic Eliashberg equations lead to overestimated values of $T_C$ (ThH$_{10}$) of 205-228 K at 200 GPa ($\mu^*$=0.15-0.1) compared to the experimental one (159-161 K). That may be explained by anharmonic character of hydrogen vibrations and increased over 0.15 values of Coloumb pseudopotential ($\mu^*$) in accordance with previously published results for H$_3$S (Ref. [13], $\Delta T_{anh}$ = -56 K), YH$_6$ (Ref. [14], $\Delta T_{anh}$ = -34 K]), LaH$_{10}$ (Refs. [15,16], $T_C$(exp)-$T_C$(theory) = 250-286 K = -36 K] and LaH$_{16}$ (Ref. [17], SCDFT yields $\mu^*$ over 0.21). Results of calculation of $T_C$ using Allen-Dynes (A-D) formula leads to reasonable values of $T_C$(ThH$_{10}$) = 150-183 K, that is why we preferred to use A-D formula instead of full Eliashberg equations.

To calculate the isotopic coefficient $\beta$ the Allen-Dynes interpolation formulas were used:

$$\beta_{McM} = -\frac{dlnT_C}{dlnM} = \frac{1}{2}\left[1 - \frac{1.04(1+\lambda)(1+0.62\lambda)}{[\lambda - \mu^*(1+0.62\lambda)]^2}\mu^{*2}\right] \tag{S7}$$

$$\beta_{AD} = \beta_{McM} - \frac{2.34\mu^{*2}\lambda^{3/2}}{(2.46+9.25\mu^*)\cdot((2.46+9.25\mu^*)^{3/2} + \lambda^{3/2})} - \frac{130.4\cdot\mu^{*2}\lambda^2(1+6.3\mu^*)\left(1-\frac{\omega_{log}}{\omega_2}\right)\frac{\omega_{log}}{\omega_2}}{\left(8.28+104\mu^*+329\mu^{*2}+2.5\cdot\lambda^2\frac{\omega_{log}}{\omega_2}\right)\cdot\left(8.28+104\mu^*+329\mu^{*2}+2.5\cdot\lambda^2\left(\frac{\omega_{log}}{\omega_2}\right)^2\right)} \tag{S8}$$

where the last two correction terms are usually small (~0.01).

The Sommerfeld constant was found as

$$\gamma = \frac{2}{3}\pi^2 k_B^2 N(0)(1+\lambda)., \tag{S9}$$

and was used to estimate the upper critical magnetic field and the superconductive gap in $P6_3/mmc$-ThH$_9$ at 100 GPa and 150 GPa by well-known semi-empirical equations of the BCS theory (Ref.[18], Equations 4.1 and 5.11), working satisfactorily for $T_C/\omega_{\log} < 0.25$:

$$\frac{\gamma T_C^2}{H_C^2(0)} = 0.168\left[1 - 12.2\left(\frac{T_C}{\omega_{log}}\right)^2 \ln\left(\frac{\omega_{log}}{3T_c}\right)\right] \tag{S10}$$

$$\frac{2\Delta(0)}{k_B T_C} = 3.53\left[1 + 12.5\left(\frac{T_C}{\omega_{log}}\right)^2 \ln\left(\frac{\omega_{log}}{2T_c}\right)\right] \tag{S11}$$

The lower critical magnetic field was calculated according to the Ginzburg-Landau theory, [15]

$$\frac{H_{C1}}{H_{C2}} = \frac{\ln k}{2\sqrt{2}k^2}, k = \lambda_L/\xi \tag{S12}$$

where $\lambda_L$ is the London penetration depth, found in SI as

$$\lambda_L = 1.0541 \cdot 10^{-5}\sqrt{\frac{m_e c^2}{4\pi n_e e^2}} \tag{S13}$$

where $c$ is the speed of light, $e$ is the electron charge, $m_e$ is the mass of an electron, and $n_e$ is an effective concentration of charge carriers, expressed from the average Fermi velocity ($V_F$) in the Fermi-gas model:

$$n_e = \frac{1}{e\pi^2}\left(\frac{m_e V_F}{\hbar}\right)^3 \tag{S14}$$

The coherence length $\xi$ was found as $\xi = \sqrt{\hbar/2e(\mu_0 H_{C2})}$ and was used to estimate the average Fermi velocity

$$V_F = \frac{\pi \cdot \Delta(0)}{\hbar}\xi \tag{S15}$$

To figure out the upper critical magnetic field $\mu_0 H_{c2}(0)$, the extrapolation method combined with the Ginzburg-Landau (GL) equation

$$\mu_0 H_{C2}(T) = \mu_0 H_{C2}(0)\left(1 - \frac{T^2}{T_C^2}\right) \tag{S16}$$

was applied. Werthamer-Helfand-Hohenberg (WHH) model [19] for critical magnetic field, simplified by Baumgartner et al. [20]:

$$\mu_0 H_{C2}(T) = \frac{\mu_0 H_{C2}(0)}{0.693}\left(\left(1-\frac{T}{T_C}\right) - 0.153\cdot\left(1-\frac{T}{T_C}\right)^2 - 0.152\cdot\left(1-\frac{T}{T_C}\right)^4\right) \tag{S17}$$